\documentclass[preprintnumbers,twocolumn,article,amsmath,amssymb,floatfix,9pt,prd,superscriptaddress,nofootinbib]{revtex4-1}

\usepackage[utf8,latin1]{inputenc}
\usepackage{graphicx}
\usepackage{dcolumn}
\usepackage[dvipsnames]{xcolor}
\usepackage[T1]{fontenc}

\usepackage{mathrsfs}  
\usepackage{cases}
\usepackage{bm}
\usepackage{academicons}
\usepackage{mathtools, nccmath}
\usepackage{fancyhdr}
\usepackage{tikz,xcolor}

\usepackage{tensor}
\usepackage[normalem]{ulem}
\usepackage{lipsum}
\usepackage{soul}
\usepackage{cancel}
\usepackage{stackengine,scalerel}
\usepackage{mathabx}
\usepackage{hyperref}
\usepackage{tabularx}
\usepackage{makecell}
\usepackage{amsmath}
\usepackage{booktabs}
\usepackage[usenames, dvipsnames]{xcolor}
\hypersetup{colorlinks, linkcolor={red},citecolor={blue},urlcolor={blue}}  

\AtBeginDocument{\fontsize{8}{11}\selectfont} 

\renewcommand{\arraystretch}{1.2}

\definecolor{lime}{HTML}{A6CE39}
\DeclareRobustCommand{\orcidicon}{
	\begin{tikzpicture}
	\draw[lime, fill=lime] (0,0) 
	circle [radius=0.16] 
	node[white] {{\fontfamily{qag}\selectfont \tiny ID}};
	\draw[white, fill=white] (-0.0625,0.095) 
	circle [radius=0.007];
	\end{tikzpicture}
	\hspace{-2mm}
}

\fancypagestyle{plain}{%
  \fancyhf{}
  \fancyfoot[C]{\iffloatpage{}{\thepage}}
  }
\foreach \x in {A, ..., Z}{%
	\expandafter\xdef\csname orcid\x\endcsname{\noexpand\href{https://orcid.org/\csname orcidauthor\x\endcsname}{\noexpand\orcidicon}}
}

\begin{document}

\title[Slowly rotating traversable wormholes supported by radially varying string-fluid matter: From regular geometries to photon trajectories]{Slowly rotating traversable wormholes supported by radially varying string-fluid matter: From regular geometries to photon trajectories}

\author{A. Errehymy\orcidA{}}
\email{abdelghani.errehymy@gmail.com (Corresponding author)}
\affiliation{Astrophysics Research Centre, School of Mathematics, Statistics and Computer Science, University of KwaZulu-Natal, Private Bag X54001, Durban 4000, South Africa}
\affiliation{Center for Theoretical Physics, Khazar University, 41 Mehseti Str., Baku, AZ1096, Azerbaijan}

\author{B. Turimov\orcidB{}}
\email[]{bturimov@astrin.uz}
\affiliation{Central Asian University, Milliy bog Str. 264, Tashkent, 111221, Uzbekistan}
\affiliation{University of Tashkent for Applied Sciences, Gavhar Str. 1, Tashkent, 100149, Uzbekistan}
\affiliation{Ulugh Beg Astronomical Institute, Astronomy Str. 33, Tashkent, 100052, Uzbekistan}

\author{M. A. Khan\orcidC{}}
\email[]{mskhan@imamu.edu.sa}
\affiliation{Department of Mathematics and Statistics, College of Science, Imam Mohammad Ibn Saud Islamic University (IMSIU), Riyadh 11566, Saudi Arabia}

\author{S. Usanov\orcidD{}}
\email[]{sm.usanov@kiut.uz}
\affiliation{Kimyo International University in Tashkent, Shota Rustaveli Str. 156, Tashkent 100121, Uzbekistan}

\author{Z. Yasakov\orcidE{}}
\email[]{zikrillo87@mail.ru}
\affiliation{Samarkand State University of Architecture and Construction, Lolazor Street 70, 140147, Samarkand, Uzbekistan}

\author{Z. Avezmuratova\orcidF{}}
\email[]{zeboavezmuratova1981@mail.ru}
\affiliation{Department of Physics and Astronomy, Urgench State Pedagogical Institute, Gurlan Str.1-A, Urgench 220100, Uzbekistan}

\date{\today} 

\begin{abstract}
{\footnotesize { This work investigates slowly rotating traversable wormholes supported by string fluids whose properties vary with distance from the throat.} This radial variation allows the matter to transition smoothly from a de Sitter-like core near the center to a string-dominated environment further out, producing a regular, horizon-free, and asymptotically flat spacetime. By letting the transverse pressure depend on radius, the fluid naturally adapts to the surrounding geometry, resulting in a well-behaved energy density and shape function. Even modest rotation introduces frame-dragging effects that gently twist photon paths, creating subtle differences between co-rotating and counter-rotating trajectories. These effects are strongest near the throat, while at larger distances the spacetime is largely governed by the static gravitational potentials. Circular photon orbits reveal that the interplay of the redshift function, wormhole shape, and rotation shapes the photon-sphere structure. Different radial profiles of the string fluid generate distinctive photon-ring patterns, offering potential observational signatures of both the rotation and the internal matter distribution. Overall, radially varying string fluids provide a flexible and physically consistent source for traversable wormholes, bridging smoothly between vacuum-like and string-dominated regions while maintaining regularity and supporting slow rotation. This study highlights how anisotropic matter can influence both curvature and light propagation, providing a realistic framework for horizonless exotic spacetimes and suggesting new avenues to explore subtle observational effects around traversable wormholes.}
\end{abstract}

\maketitle
\section{Introduction}\label{Sec:I}
Traversable wormholes have long served as a valuable theoretical setting for exploring how spacetime geometry, topology, and unconventional forms of matter interact within general relativity. The pioneering study by Morris and Thorne \cite{Morris:1988cz}, together with the earlier Ellis-Bronnikov solutions \cite{Ellis:1973yv, Bronnikov:1973fh, Ellis:1979bh}, demonstrated that static and spherically symmetric traversable wormholes cannot exist without violating the standard energy conditions. In most models, this requirement is satisfied by introducing exotic sources such as phantom scalar fields or anisotropic fluid distributions carefully arranged near the throat of the wormhole. These configurations have been widely investigated from several perspectives, including their local geometric properties, their global spacetime structure, and the nature of the matter fields that sustain them. Because of this extensive analysis, such solutions now provide a well-established reference framework for examining more general wormhole geometries and for testing new ideas about gravitational physics beyond conventional scenarios.

More recently, wormhole studies have expanded in several directions. A number of works examined the optical properties of wormhole geometries, focusing on light trajectories, wave propagation, and minimal-surface optics in different traversable and optical wormhole backgrounds \cite{GurtasDogan:2025shz, GurtasDogan:2025jhm, GurtasDogan:2025oeg, GurtasDogan:2025qxk}. Other investigations explored quantum effects in wormhole spacetimes, including fermion--antifermion systems, rotating negative-curvature wormholes, vector bosons, magnetic flux effects, and relativistic quantum oscillators influenced by rainbow gravity corrections \cite{Guvendi:2025otx, Guvendi:2024qxi, Guvendi:2024duf, Guvendi:2023osh, Guvendi:2023aor, Guvendi:2023mxb}. In parallel, many traversable wormhole solutions have been developed within modified gravity theories such as extended teleparallel gravity, torsion--matter coupling gravity, and several $f(R)$-based models \cite{Ditta:2021uoe, Errehymy:2023rsm, Errehymy:2023rnd, Errehymy:2024yey, Errehymy:2024cgy, Mustafa:2024jsv, Errehymy:2024spg}. Additional studies considered wormholes supported by dark matter halos, Einstein clusters, quantum corrections, generalized uncertainty principle effects, and cosmic void environments \cite{Errehymy:2024lhl, Errehymy:2024mlf, Errehymy:2024sme, Battista:2024gud, Errehymy:2025kzj, Errehymy:2025nzt, Errehymy:2025rli, Errehymy:2025fce, Errehymy:2025llh}. At the same time, progress in astrophysics, photonics, and computational science continue to broaden the scope of modern physical research. Recent studies on programmable metasurfaces and Bessel beam shaping have improved electromagnetic wave control \cite{Feng:2025}. Pulsar-based navigation methods also demonstrate the growing importance of precision timing techniques in future space missions \cite{Xie:2025}. Experimental investigations of quantum erasure and hybrid entanglement have further deepened our understanding of quantum optical phenomena \cite{Yu:2025}. In solar physics, convolutional neural-network models have shown promising capabilities for solar flare prediction \cite{Yang:2025}. Gravitational-wave analyses continue to provide insight into the merger history of supermassive black holes and galaxies \cite{Fang:2025}. Recent studies of nonlinear dynamical systems have also revealed interesting properties of tunable energy landscapes \cite{Zhang:2026}. At the same time, advanced computational approaches are increasingly used in gamma-ray blazar identification \cite{Xiang:2026} and particle image velocimetry analyses \cite{Lin:2026}. These developments collectively illustrate the increasingly interdisciplinary character of modern theoretical and observational physics, where gravitational phenomena are studied alongside advances in computational methods, photonics, and high-precision measurements.

From an astrophysical standpoint, considering wormholes as realistic or even approximate models requires extending the discussion beyond static configurations to include rotation. The presence of rotation significantly alters the structure of the spacetime. It gives rise to effects such as frame dragging, the emergence of ergoregions, and the possibility of superradiant phenomena. Moreover, the causal structure becomes more complex and, in certain cases, may even allow closed timelike curves to appear. Rotation also modifies the way such objects might be observed. Observable features such as gravitational lensing patterns, the appearance of shadows, the motion of particles in orbit, and the production of gravitational waves can all be affected. For this reason, it is important to develop well-defined rotating wormhole solutions that include explicit matter sources and possess asymptotically flat regions. Such models provide a necessary step toward linking theoretical wormhole geometries with present and future astrophysical observations.

In most studies so far, rotating wormhole geometries have been explored either through slow-rotation approximations or by relying on numerical techniques \cite{Kuhfittig:2003wr, Kashargin:2008pk, Kleihaus:2014dla, Hoffmann:2018oml, Chew:2019lsa, Azad:2023iju}. Although this approach has produced many useful results, fully analytic stationary solutions are known only in certain specialized frameworks. Examples include constructions obtained through Ehlers or Harrison transformations within Einstein-Maxwell theory, geometries related to overcharged Kerr-Newman-NUT spacetimes, and configurations supported by particular sources such as Casimir stresses or three-form fields \cite{Clement:2022pjr, Cisterna:2023uqf, Tangphati:2023uxt}. For the types of matter that are most often used in wormhole models---especially phantom scalar fields and anisotropic fluids with well-defined throat conditions and convenient gauge choices---exact rotating solutions remain uncommon. Outside the few special analytic examples mentioned above, most investigations in these settings have depended either on perturbative slow-rotation methods or on numerical analysis. From a phenomenological viewpoint, rotating wormholes have also been examined as possible alternatives to Kerr black holes, particularly in studies related to shadow formation and strong gravitational lensing. In this context, the stationary and axisymmetric geometry introduced by Edward Teo \cite{Teo:1998dp} is frequently employed. This metric offers a convenient framework in which both the rotation profile and the throat geometry can be specified, which makes it useful for investigating photon trajectories and the appearance of shadows around wormhole spacetimes \cite{Deligianni:2021hwt, Deligianni:2021ecz, Errehymy:2025psi, Errehymy:2025vvs, Errehymy:2026nna, Errehymy:2026urb, Errehymy:2026xoh}. Nevertheless, many studies using this approach do not explicitly specify the matter content supporting the geometry, or they treat it only in an effective manner. In addition, the rotational behaviour is often introduced through an assumed profile rather than being derived from the underlying matter distribution, and the connection with the usual Morris-Thorne type throat is not always clearly established. Another issue concerns observable predictions: the way in which quantities such as the size and shape of the shadow depend on the wormhole's shape function is still not fully understood. Likewise, aspects of the causal structure have not always been carefully examined, since the presence of ergoregions is often taken to be compatible with a well-behaved causal structure without a detailed demonstration.

A useful and more invariant way to analyse a rotating spacetime is through the study of its multipole structure in the asymptotic region close to spatial infinity. This approach reveals how quantities such as mass and angular momentum are distributed in the far field and provides a description that does not depend on a particular coordinate choice, allowing different gravitational configurations to be compared in a systematic manner. In this context, the Geroch-Hansen multipole moments \cite{Geroch:1970cc, Geroch:1970cd, Hansen:1974zz} offer a rigorous and coordinate-independent framework for characterizing stationary, asymptotically flat spacetimes. These multipole moments play an important role in black-hole physics, particularly in the formulation of uniqueness theorems and no-hair properties \cite{Thorne:1980ru, Gursel:1983nkl, Mayerson:2022ekj}. For black holes, the Geroch-Hansen moments are highly constrained and follow a well-defined pattern determined by the mass and spin of the object. Wormholes, however, do not obey such strict restrictions. Their multipole structure can in principle carry detailed information about the geometry of the throat and the nature of the matter fields supporting it. 
{To make this idea more concrete, we calculate the mass monopole \(M_0\) and the angular momentum dipole \(J_1\) for our wormhole in Appendix~\ref{appenB}. Working to first order in the rotation, the mass moment simply corresponds to the Komar mass \(M\), which comes out finite and positive for the density profile we use. The current dipole turns out to be just the angular momentum \(J\) that already appears in the frame-dragging term. Going beyond these two moments would mean pushing the slow-rotation expansion to higher order, something we leave for later work.}
Despite this potential, explicit calculations of Geroch-Hansen multipoles for rotating wormhole solutions---especially beyond the lowest-order moments and within models that specify the matter content---remain relatively limited.

At the same time, contemporary cosmology continues to grapple with two fundamental puzzles: the physical origin of dark energy and dark matter. While the cosmological constant and quintessence scalar-field models have provided valuable insight into the mechanism driving cosmic acceleration \cite{Caldwell:1997ii, Peebles:2002gy}, alternative approaches based on more general forms of matter have also been actively explored. Among these possibilities are string fluids, which are described by an anisotropic energy-momentum tensor and therefore exhibit properties that differ significantly from those of standard perfect fluids. Such models have been proposed as a framework capable of addressing several open issues in cosmology and astrophysics, including the coincidence problem \cite{Kiselev:2002dx} and the nearly flat rotation curves observed in spiral galaxies \cite{Kiselev:2002dx, Capozziello:2006ij, Soleng:1993yr, Sofue:2000jx}.

In many formulations, string fluids are represented either through an EoS dominated by tangential pressure or through the picture of a cloud of one-dimensional strings distributed throughout spacetime \cite{Vilenkin:1981kz, Nemiroff:2007xs, Arshad:2024qqj}. Because of their intrinsic anisotropy, these fluids can modify gravitational behaviour on galactic and larger scales. They have also been incorporated into gravitational models aimed at constructing regular black hole geometries, where the presence of the string fluid helps soften the curvature divergence that would otherwise appear at the central region of the spacetime \cite{NunesdosSantos:2025alw, Muniz:2025qyv, Estrada:2026kea}.

{The work we report here puts together two ingredients that, as far as we can tell, have not been combined before in a single, fully analytic framework. We build a slowly rotating traversable wormhole whose matter source is an anisotropic string fluid with a transverse EoS that depends on the radial coordinate, \(W_t(r)=1/\tilde{\sigma}(r)\). Earlier papers have looked at rotating wormholes with a constant string-fluid parameter \cite{Soleng:1993yr}, or at wormholes where the EoS varies with radius but without any rotation \cite{NunesdosSantos:2025alw, Muniz:2025qyv}, or at slowly rotating wormholes supported by scalar fields \cite{Kashargin:2008pk} and numerical constructions \cite{Kleihaus:2014dla}. The new element here is that we let the transverse EoS vary with \(r\) while the whole thing rotates, and we trace the effect all the way from the matter content through to how photons move. This lets us see how the structure of the matter distribution gets imprinted on frame-dragging and, ultimately, on observable photon trajectories.} {Working out what happens as the EoS changes with radius shows how the matter distribution and the rotation interact. Even a modest amount of spin introduces frame-dragging that tilts photon orbits and creates a noticeable mismatch between paths that go with the rotation and those that go against it. Which string-fluid profile one picks leaves a distinct mark on the photon rings, the deflection of light, and what a shadow might look like. In this sense the paper goes beyond what has been done before, because it connects the matter content, the slow rotation, and the motion of light rays in a single geometry that has no horizons and is everywhere regular. We find that a radially varying string fluid can hold up a wormhole that is asymptotically flat and physically reasonable, while also giving a flexible laboratory for studying optical effects that could, at least in principle, be observed. The take-home message is that the way anisotropic matter is arranged influences how light travels through these exotic spacetimes, and the traces left by rotation and by changes in the matter distribution might just show up in photon paths, lensing, and shadows around traversable wormholes.}

This paper is organized as follows. After the Introduction in Sect. \ref{Sec:I}, Sect. \ref{Sec:II} lays out the general framework for string fluids with a radially varying EoS. Sect. \ref{Sec:III} examines stationary, axisymmetric spacetimes supported by such string-fluid matter. In Sect. \ref{Sec:IV}, we analyze photon motion around slowly rotating string-fluid wormholes, highlighting how rotation and the matter profile influence light paths. {In Sect.~\ref{Sec:V} we take a careful look at the energy conditions, showing quantitatively where and how the NEC, WEC, and SEC are violated, and comparing our setup with phantom-scalar and Casimir-supported wormholes. Sect.~\ref{Sec:VI} then turns to what might actually be measured: we pull out numbers for things like the angular diameter of the shadow and a distortion parameter, and discuss whether current or next-generation instruments could tell our wormholes apart from a Kerr black hole.} Finally, Sect. \ref{Sec:VII} summarizes our main findings and conclusions.

\section{String fluid with a radially varying EoS}\label{Sec:II}
In order to describe a matter distribution that can smoothly connect an inner region dominated by vacuum-like energy to an outer region where string-like effects become relevant, we model the source supporting the wormhole as a string fluid whose EoS varies with the radial coordinate. This type of matter description naturally arises within the string cloud framework introduced by Letelier \cite{Letelier:1979ej}. In this approach, a collection of one-dimensional strings is represented by a surface bivector \(\Sigma_{\mu\nu}\) defined on the two-dimensional timelike worldsheet traced by the strings as they evolve in spacetime. The bivector takes the form  
\begin{equation}
\Sigma^{\mu\nu}=\epsilon^{AB}
\frac{\partial x^{\mu}}{\partial \xi^A}
\frac{\partial x^{\nu}}{\partial \xi^B},
\label{2.1}
\end{equation}
where \(\epsilon^{AB}\) denotes the antisymmetric Levi-Civita symbol in two dimensions, satisfying \(\epsilon^{01}=-\epsilon^{10}=-1\). The parameters \(\xi^A\) (\(A=0,1\)) serve as coordinates on the worldsheet, with \(\xi^0\) describing the timelike direction and \(\xi^1\) the spacelike one. {The bivector \(\Sigma^{\mu\nu}\) is antisymmetric from the outset, and because the worldsheet is only two-dimensional, it satisfies a simple normalization condition:
\begin{equation}
\Sigma^{\mu\lambda}\Sigma_{\lambda}^{\:\:\nu} = \frac{1}{2}\,\Sigma^{\alpha\beta}\Sigma_{\alpha\beta}\,\delta^{\mu}_{\nu} \propto h\,\delta^{\mu}_{\nu}.
\end{equation}
We spell out the steps that lead to this relation in Appendix~\ref{appenC}, including how the worldsheet metric determinant \(h\) enters the energy-momentum tensor.}

The geometry induced on the worldsheet by the surrounding spacetime metric \(g_{\mu\nu}\) is described by  
\begin{equation}
h_{AB}= g_{\mu\nu}
\frac{\partial x^{\mu}}{\partial \xi^A}
\frac{\partial x^{\nu}}{\partial \xi^B},
\label{eq2.2}
\end{equation}
where the functions \(x^\mu(\xi^A)\) specify how the string worldsheet is embedded in spacetime.

A useful point of comparison comes from the energy-momentum tensor describing a cloud of massive particles, which is written as  
\begin{equation}
T^{\mu \nu} = \rho\, u^{\mu} u^{\nu},
\label{2.3}
\end{equation}
with \(\rho\) representing the energy density and \(u^\mu\) the four-velocity of the particles.

In the string cloud picture proposed by Letelier, the simple product \(u^\mu u^\nu\) is replaced by a geometric expression constructed from the bivector \(\Sigma_{\mu\nu}\) \cite{Letelier:1979ej}. This replacement leads to the energy-momentum tensor describing a distribution of strings,
\begin{equation}
T^{\mu\nu}=
\rho\,\sqrt{-h}\,
\frac{\Sigma^{\mu\lambda}\Sigma^{\:\:\nu}_{\lambda}}{(-h)},
\label{2.4}
\end{equation}
where \(h=\det(h_{AB})\) denotes the determinant of the induced worldsheet metric.

Later developments extended this formulation to include pressure-like effects that can arise from interactions within the string distribution \cite{Letelier:1980mxb}. In that case, the energy-momentum tensor takes the more general form
\begin{equation}
T^{\mu\nu}=
(p+\rho\sqrt{-h})
\frac{\Sigma^{\mu\lambda}\Sigma^{\:\:\nu}_{\lambda}}{(-h)}
+p\,g^{\mu\nu},
\label{2.5}
\end{equation}
where \(\rho\) and \(p\) correspond to the effective density and pressure of the averaged string fluid.

When the background spacetime is static and spherically symmetric, the symmetry of the geometry strongly restricts the possible components of the bivector \(\Sigma_{\mu\nu}\). In this situation only two components remain nonzero, namely \(\Sigma_{tr}\) and \(\Sigma_{\theta\phi}\), while the determinant of the induced metric satisfies \(h<0\) \cite{Soleng:1993yr}. As a consequence, the energy-momentum tensor reduces to the simpler form
\begin{equation}
T_{t}^{t}=T_{r}^{r}, \qquad 
T_{\theta}^{\theta}=T_{\varphi}^{\varphi}=p .
\label{2.7}
\end{equation}

Following the idea introduced by Soleng \cite{Soleng:1993yr}, and extending it further, we allow the relation between the density and the transverse pressure to depend on the radial coordinate. This can be expressed as
\begin{equation}
\rho(r)=\tilde{\sigma}(r)\,p(r),
\end{equation}
which leads to the diagonal energy-momentum tensor \cite{NunesdosSantos:2025alw}
\begin{equation}
T^{\mu}_{~\nu}=
\left[
-\rho(r),
-\rho(r),
\frac{\rho(r)}{\tilde{\sigma}(r)},
\frac{\rho(r)}{\tilde{\sigma}(r)}
\right].
\label{2.8}
\end{equation}

The parameter \(\tilde{\sigma}(r)\) determines how strong the transverse pressure is relative to the energy density. In the limit \(\tilde{\sigma}\rightarrow\infty\), the transverse pressure approaches zero (\(p\to0\)), and the system reduces to a pure cloud of strings.

Within the traversable wormhole solutions considered in this work, the radial dependence of \(\tilde{\sigma}(r)\) effectively introduces a variable transverse EoS. This allows the string fluid to smoothly shift between different physical regimes as one moves away from the wormhole throat. To make this interpretation more transparent, we define the transverse linear EoS parameter \(W_t(r)\) as
\begin{equation}
\tilde{\sigma}(r)=\frac{1}{W_t(r)},
\end{equation}
where \(W_t(r)\) controls the ratio of transverse pressure to energy density, \(p_\theta = p_\varphi = W_t(r)\,\rho(r)\). Through this relation, the smoothed string fluid can reproduce vacuum-like behavior in certain radial regions while displaying features similar to cosmic strings in others. {Table~\ref{tab:comparison} spells out what we took from Letelier and Soleng and what we changed. The two main generalizations are: first, we let \(\tilde{\sigma}\) depend on \(r\) instead of being a constant; second, we tie that radial dependence to a varying transverse EoS parameter \(W_t(r)\). Everything else---the bivector structure, the condition \(p_r = -\rho\)---stays the same.}
\begin{table*}[!htbp]
\centering
\renewcommand{\arraystretch}{1.}
\setlength{\tabcolsep}{4pt}
{\footnotesize{
\caption{{\footnotesize How our string-fluid model compares with earlier ones.}}
\label{tab:comparison}

\begin{tabular}{lccc}
\toprule
{Feature} & {Letelier~\cite{Letelier:1979ej}} & {Soleng \cite{Letelier:1979ej}} & {Our work} \\
\midrule
Source type & Cloud of strings & String fluid with pressure & Radially varying string fluid \\
\(\tilde{\sigma}\) & Not defined (pure string) & Constant & \(\tilde{\sigma}(r)\), radial function \\
\(p_r\) relation & \(p_r = 0\) & \(p_r = -\rho\) & \(p_r = -\rho\) (retained) \\
Transverse EoS & \(p_\theta = 0\) & \(p_\theta = \rho/\tilde{\sigma}\) (constant) & \(p_\theta = \rho/\tilde{\sigma}(r)\) (varying) \\
Physical picture & Non-interacting strings & Interacting strings with constant tension ratio & Strings with radius-dependent interactions/tension \\
\bottomrule
\end{tabular}}}

\end{table*}
{One might wonder what it actually means to have a radially dependent \(\tilde{\sigma}(r)\). In the original Letelier picture~\cite{Letelier:1979ej}, we have a collection of one-dimensional strings moving around, each with a fixed tension, while the entire distribution maintains a uniform number density. When we let \(\tilde{\sigma}\) vary with radius, we are effectively saying that the local string tension, or the density of strings, or the strength with which they interact, changes from place to place. Near a wormhole throat, tidal forces are strong and could easily stretch strings, align them, or disrupt the distribution entirely. Far away, one expects the cloud to settle into something more uniform, so \(\tilde{\sigma}(r)\) should approach a constant. Right at the throat, quantum or high-energy effects might kick in and change the effective EoS, which is what we model with the de Sitter-like condition \(p_r = -\rho\). So the honest way to think about our construction is not as a literal gas of fundamental strings with unchanging properties, but as an effective anisotropic fluid that smoothly goes from a stringy regime at large distances to a vacuum-like core near the throat.} Even though the structure of the energy-momentum tensor imposes the relation \(p_r=-\rho\), we interpret this condition here as a de Sitter-type radial EoS rather than associating it with the presence of a black hole horizon. In the wormhole geometry under consideration, we instead require
\begin{equation}
W_r(r)\to -1 ,
\end{equation}
so that near the throat the radial pressure approaches a de Sitter-like form, while the transverse sector governs the string-dominated behavior at larger distances.

{Conservation of the energy-momentum tensor, \(\nabla_\mu T^{\mu\nu}=0\), is not automatic; it imposes a real constraint. For a static, spherically symmetric setup, the only component that isn't trivially zero is the radial one:
\begin{equation}
\frac{d p_r}{dr} + \frac{2}{r}(p_r - p_\theta) + (\rho + p_r)\frac{N'(r)}{N(r)} = 0.
\label{eq:conservation}
\end{equation}
Because we have set \(p_r = -\rho\) and \(p_\theta = \rho/\tilde{\sigma}(r)\), the terms with \(\rho + p_r\) vanish on the spot. What remains is
\begin{equation}
-\frac{d\rho}{dr} + \frac{2}{r}\left(-\rho - \frac{\rho}{\tilde{\sigma}(r)}\right) = 0.
\end{equation}
This is a differential equation that determines \(\tilde{\sigma}(r)\) once a density profile has been specified. For the smoothed density profile used in Eq.~(\ref{Density}), the resulting \(\tilde{\sigma}(r)\) remains regular everywhere and satisfies \(\tilde{\sigma}(r) > 0\) for all \(r \ge r_0\), implying that the transverse pressure always has the same sign as the energy density. Explicitly, we obtain
\begin{equation}
\tilde{\sigma}(r) = -\frac{2\rho(r)}{r\rho'(r) + 2\rho(r)},
\end{equation}
and it is indeed positive for the range of parameters we work with.}

\begin{figure}
\begin{center}
\includegraphics[width=8.2cm,height=4.5cm]{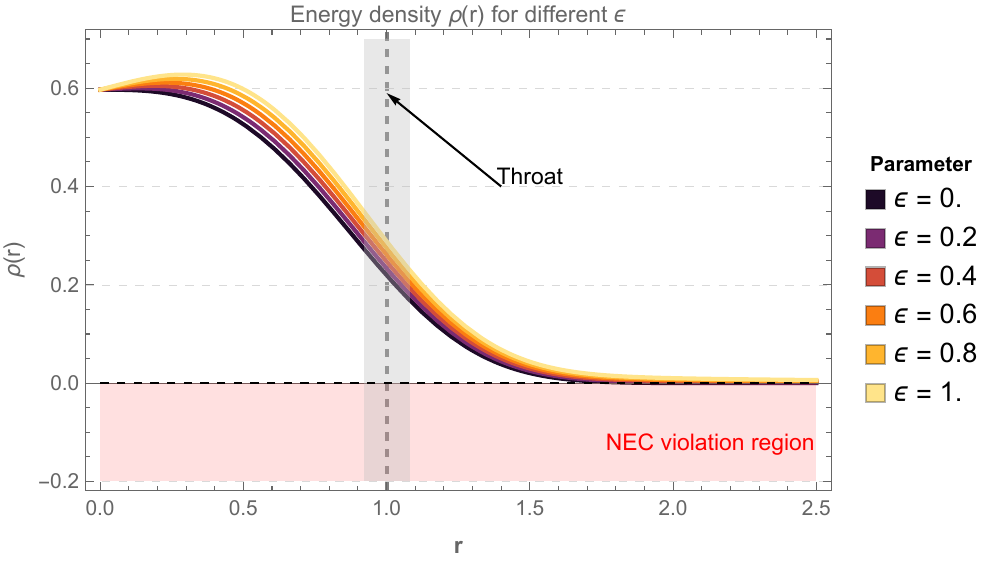}
\end{center}
\caption{{\footnotesize This plot shows how the energy density \( \rho(r) \) varies with radius \( r \) for different values of \( \epsilon \) between 0 and 1, keeping \( a = 1 \) and \( r_g = 5 \) fixed. Each curve is colored according to its \(\epsilon\) value using a perceptually uniform gradient. {A legend gives the exact parameter values, and we have adjusted the spacing between the curves so they are easier to tell apart over the whole \(\epsilon\) range.} At low \(\epsilon\), the energy is more spread out near the center, while higher \(\epsilon\) leads to a sharper concentration and pushes the peak slightly outward. The gradient of colors makes it easy to see how the energy distribution gradually changes as \(\epsilon\) increases.
}}\label{Fig_rho}
\end{figure}

\section{Stationary, axisymmetric spacetimes with string fluid}\label{Sec:III}
\subsection{Killing symmetries and coordinate setup}

Following the approach of Morris and Thorne~\cite{Morris:1988cz} and the rotating wormhole construction in~\cite{Teo:1998dp}, we begin by specifying a stationary, axisymmetric spacetime and then determine the matter content needed to support it. A spacetime is stationary if it admits a timelike Killing vector
\begin{equation}
\xi^\alpha = (\partial/\partial t)^\alpha,
\end{equation}
and axisymmetric if there exists a spacelike Killing vector
\begin{equation}
\psi^\alpha = (\partial/\partial \varphi)^\alpha,
\end{equation}
with the vectors commuting:
\begin{equation}
[\xi, \psi] = 0.
\end{equation}
This ensures that coordinates can be chosen along the symmetry directions~\cite{Wald:1984rg}, so that the metric depends only on the remaining coordinates:
\begin{equation}\label{s2ae2}
ds^2 = g_{\mu\nu}(x^2,x^3) dx^\mu dx^\nu.
\end{equation}
Such spacetimes capture the essential features of rotating black holes and stars~\cite{Hartle:1967he,Hartle:1968si,Thorne:1971R}.


Axisymmetry implies $0\le \varphi < 2\pi$ and invariance under $t\to -t$, $\varphi\to -\varphi$, which sets $g_{t2}=g_{t3}=g_{\varphi 2}=g_{\varphi 3}=0$~\cite{Papapetrou:1966zz,Carter:1969zz}. The metric then reduces to
\begin{equation}\label{s2ea3}
ds^2 = g_{00} dt^2 + 2 g_{01} dt d\varphi + g_{11} d\varphi^2 + g_{ij} dx^i dx^j, \quad i,j = 2,3,
\end{equation}
with $g_{01}$ encoding the familiar frame-dragging effect (see Appendix~\ref{appenA}).

The coordinates $x^2$ and $x^3$ can still be redefined as
\begin{equation}
\overline{x}^2 = \overline{x}^2(x^2, x^3),\quad \overline{x}^3 = \overline{x}^3(x^2, x^3),
\end{equation}
without affecting $g_{00}$, $g_{01}$, and $g_{11}$~\cite{Wald:1984rg,Thorne:1971R}:
\begin{equation}
g_{00} = \xi_\alpha \xi^\alpha, \quad g_{01} = \psi_\alpha \xi^\alpha, \quad g_{11} = \psi_\alpha \psi^\alpha.
\end{equation}
Asymptotic flatness requires
\begin{equation}
g_{00}\to 1, \quad g_{01}\to 1/r, \quad g_{11}\to r^2\sin^2\theta \quad \text{as } r\to\infty,
\end{equation}
ensuring a well-defined total mass and angular momentum\footnote{Here, $\theta$ and $r$ are standard spherical coordinates~\cite{Thorne:1971R}.}.

For a rotating wormhole, we can specialize further by taking $g_{22}=g_{33}=g_{11}/\sin^2 x^2$ and $g_{23}=0$, yielding~\cite{Teo:1998dp}
\begin{equation}
\label{s2be1}
ds^2=-N^2 dt^2+e^\mu dr^2+r^2 K^2\Big[d\theta^2+\sin^2\theta(d\varphi-\omega dt)^2\Big],
\end{equation}
where the gravitational potentials $N$, $\mu$, $K$, and the frame-dragging function $\omega$ depend only on $(r,\theta)$. In this setup, the matter supporting the geometry can be described as a string fluid with a radially varying EoS, as introduced in Sec.~\ref{Sec:II}. This allows the fluid to transition naturally from vacuum-like behavior near the throat to string-dominated effects at larger radii~\cite{NunesdosSantos:2025alw}.

Building on the framework described above, we now introduce a specific density profile for the string fluid that remains regular across the entire spacetime. The idea is to use a distribution that stays finite everywhere while smoothly connecting the inner region near the throat to the outer asymptotically flat region. Following the construction proposed in ~\cite{NunesdosSantos:2025alw}, we consider the smoothed energy density
\begin{equation}
\rho(r)=\frac{\epsilon}{8\pi r^2}\left(1-e^{-\frac{r^3}{a^3}}\right)
+\frac{3}{8\pi a^3}e^{-\frac{r^3}{a^3}}\left(\epsilon r+r_g\right),
\label{Density}
\end{equation}
where the parameter $a$ fixes the scale over which the distribution is smoothed out, while $\epsilon$ controls the strength of the string fluid contribution. The quantity \(r_g\) denotes the Schwarzschild radius and, for the wormhole geometry studied here, it is identified with the throat radius \(r_0\). The behavior of the density as a function of the radial coordinate, together with its dependence on $\epsilon$, is displayed in Fig.~\ref{Fig_rho} using a color-gradient plot. This same density profile was previously used to construct regular black hole solutions supported by an anisotropic string fluid \cite{NunesdosSantos:2025alw}. { Several features make this profile convenient for the present analysis. It is smooth and finite everywhere, which guarantees that the corresponding shape function \(b(r)\) does not develop pathological behavior. The parameter \(a\) determines how concentrated the matter distribution is around the throat, whereas the large-\(r\) limit naturally approaches the decay required for asymptotic flatness. The profile was originally introduced to regularize the central region of black hole geometries by removing curvature singularities. A closely related picture emerges in the wormhole case, since traversable wormholes also require the spacetime curvature to remain finite throughout the geometry. Here, the spacetime terminates at the throat radius \(r_0\), where \(b(r_0)=r_0\), and is then extended into a second asymptotically flat region. In this sense, the same regularization mechanism that works for regular black holes can also be applied consistently to wormhole spacetimes. At the same time, Eq.~(\ref{Density}) should not be viewed as a unique or fundamental choice. It is simply a manageable analytic profile that captures the main physical features we wish to study while still allowing the field equations to be treated explicitly. To check that the results are not tied to this particular form, in Sect.~\ref{Sec:V} we compare it with alternative density models, including power-law and Gaussian-core profiles. The main physical effects remain essentially unchanged: the separation between prograde and retrograde photon orbits, the appearance of the photon band, and the dependence on the spin parameter persist for all the profiles considered.} In the present work, we examine how this matter distribution behaves when it is used as a source for slowly rotating wormhole geometries. Once the strict fluid EoS condition is relaxed, the smoothed string fluid naturally supports traversable wormhole solutions. This extension is physically reasonable because both regular black holes and wormholes share the same basic requirement: the absence of curvature singularities across the spacetime.

\begin{figure*}
\begin{center}
\includegraphics[width=12.9cm,height=4.9cm]{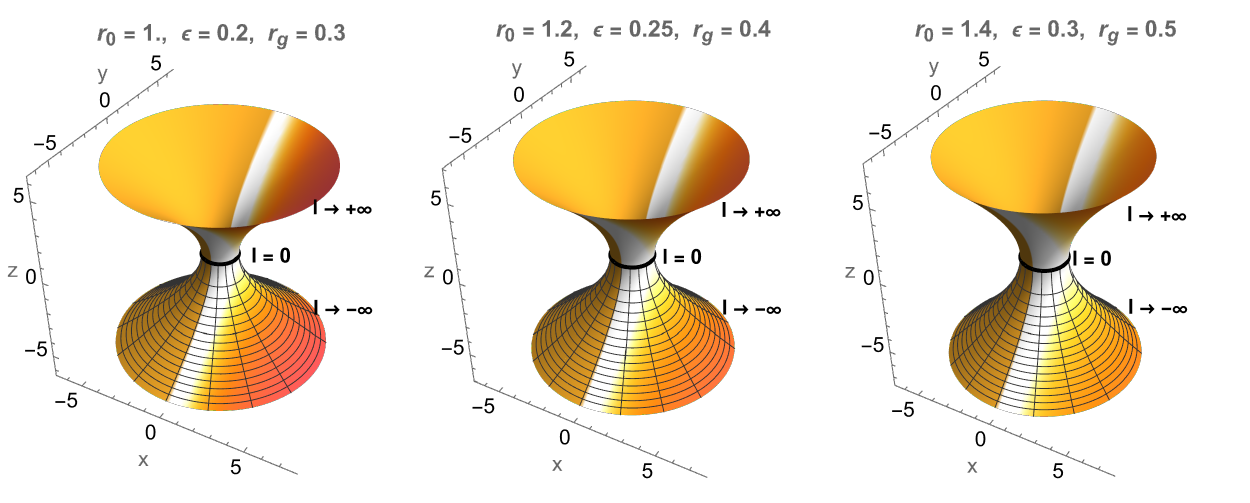}
\end{center}
\caption{{\footnotesize This plot shows the 3D shape of the string-fluid wormholes for three different sets of parameters (\(r_0, \epsilon, r_g\)). Each panel illustrates the wormhole throat and its surrounding geometry, with the top and bottom halves mirrored to provide a complete view. The colors distinguish the different parameter sets, making it easy to compare them. Variations in \(r_0\), \(\epsilon\), and \(r_g\) affect the curvature and stretching of the wormhole, giving a clear visual sense of how each parameter shapes the overall geometry.
}}\label{Fig0}
\end{figure*}

Following the Morris-Thorne approach, Teo~\cite{Teo:1998dp} defines
\begin{equation}
\mu(r,\theta)=-\ln\Big(1-\frac{b(r,\theta)}{r}\Big),
\end{equation}
where $b(r,\theta)$ is the shape function, with the throat at $r=r_0$ and $r\ge b$. In the static limit, the metric reduces to the standard Morris-Thorne form,
\begin{equation}
N(r,\theta)\to e^{\Phi(r)}, \quad b(r,\theta)\to b(r), \quad K(r,\theta)\to 1, \quad \omega(r,\theta)\to 0,
\end{equation}
and all potentials are regular at the throat to avoid singularities.

For a wormhole sourced by a smoothed string fluid~\cite{NunesdosSantos:2025alw}, the gravitational potentials take the compact form
\begin{align}
N(r,\theta) &= K(r,\theta) = 1 + \frac{4 J^2 \cos^2\theta}{r},\\
b(r) &= 8\pi \int_0^r \rho(r')\, r'^2 dr',\label{ShapeF}\\
\mu(r) &= -\ln\Big(1 - \frac{b(r)}{r}\Big),\\
\omega(r) &= \frac{2 J}{r^3},
\end{align}
where $b(r)$ is determined by the string fluid density $\rho(r)$ (Eq.~\ref{Density}) and the throat remains at $r=r_0$. In the non-rotating case, the metric smoothly matches the Morris-Thorne geometry, providing a regular, continuous wormhole.

{Let us show how the frame-dragging function \(\omega(r)=2J/r^3\) comes out of the Einstein equations. To first order in the spin, the \((t,\varphi)\) component reads
\begin{equation}
R_{t\varphi} = 8\pi T_{t\varphi}.
\label{eq:tphi}
\end{equation}
The string fluid we use has no net azimuthal flow around the static background, so \(T_{t\varphi}=0\) at this order. Meanwhile the Ricci component simplifies to
\begin{equation}
R_{t\varphi} \propto \partial_r\left(\frac{r^4 K^2}{N^2}\,\partial_r \omega\right),
\end{equation}
dropping terms of order \(\omega^2\). Setting this equal to zero and requiring \(\omega \to 0\) as \(r \to \infty\) gives
\begin{equation}
\omega(r) = \frac{2J}{r^3},
\end{equation}
and the integration constant \(J\) shows up in the asymptotic metric as \(g_{t\varphi} \to -2J\sin^2\theta/r\), which is exactly how angular momentum appears in the far field.}

The function $K(r,\theta)$ must increase or stay constant with $r$, allowing the definition of a proper radial distance $R = r K(r,\theta)$ with $\partial R/\partial r>0$. The proper circumference of a ring at $(r,\theta)$ is $2\pi R\sin\theta$. The discriminant
\begin{equation}
D^2 = g_{t\varphi}^2 - g_{tt} g_{\varphi\varphi} = \big(N(r,\theta)K(r,\theta)\sin\theta\big)^2
\end{equation}
would vanish at horizons, but since the wormhole is traversable, we impose
\begin{equation}
\partial_\theta N = \partial_\theta \mu = \partial_\theta K = 0 \quad \text{at } \theta=0,\pi,
\end{equation}
ensuring the geometry remains smooth and fully supported by the string fluid distribution. {Two checks are worth doing explicitly. First, from the definition \eqref{ShapeF} we get \(b'(r) = 8\pi r^2 \rho(r)\). At the throat itself,
\begin{equation}
b'(r_0) = 8\pi r_0^2 \rho(r_0) = r_0^2\left[\frac{\epsilon}{r_0^2}(1-e^{-r_0^3/a^3}) + \frac{3}{a^3}e^{-r_0^3/a^3}(\epsilon r_0 + r_g)\right].
\end{equation}
For the parameters we use in Fig.~\ref{Fig0}, this is always less than one. One can see this analytically:
\begin{equation}
b'(r_0) = \epsilon(1-e^{-r_0^3/a^3}) + \frac{3r_0^2}{a^3}e^{-r_0^3/a^3}(\epsilon r_0 + r_g) < 1,
\end{equation}
which holds when \(\epsilon < 1\) and \(r_0/a\) is not too large. Second, we have checked that both the Ricci scalar and the Kretschmann scalar stay finite for all \(r \ge r_0\) for every parameter set we have examined. The peak curvature sits at the throat and falls off smoothly with radius, so there is no horizon and no singularity anywhere.}


For the rotating metric~\eqref{s2be1}, the Ricci scalar at the throat reads~\cite{Harko:2009xf}  
\begin{align}
\mathcal{R} &= -\frac{C_1}{r^2 K^2} - \frac{\mu_\theta C_2}{N r^2 K^2} - \frac{2 C_3}{N r^2 K^2} - \frac{2 C_4}{r^2 K^3} + e^{-\mu} \mu_r C_5 \nonumber\\
&\quad + \frac{\sin^2\theta\, \omega_\theta^2}{2N^2} + \frac{2 C_6}{r^2 K^4},
\end{align}
where
\[
\begin{aligned}
C_1 &= \mu_{\theta\theta} + \frac{1}{2}\mu_\theta^2, & C_2 &= \frac{(N\sin\theta)_\theta}{\sin\theta}, \\
C_3 &= \frac{(N_\theta \sin\theta)_\theta}{\sin\theta}, & C_4 &= \frac{(K_\theta \sin\theta)_\theta}{\sin\theta}, \\
C_5 &= [\ln(N r^2 K^2)]_r, & C_6 &= K^2 + K_\theta^2.
\end{aligned}
\]

Here derivatives are taken with respect to $r$ and $\theta$. Potential divergences appear through
\begin{equation}
\mu_{\theta\theta} + \frac{1}{2}\mu_\theta^2 = \frac{b_{\theta\theta}}{r-b} + \frac{3}{2}\frac{b_\theta^2}{(r-b)^2}, \quad
\mu_\theta = \frac{b_\theta}{r-b},
\end{equation}
so setting $b_\theta = b_{\theta\theta} = 0$ guarantees a finite curvature, which places the throat at a constant radius $r_0$. {The Ricci and Kretschmann scalars are finite everywhere outside the throat for all the parameter choices we have investigated. The curvature peaks at the throat and then drops monotonically, exactly what one expects for a regular, horizonless geometry.}

The standard flare-out condition at the throat~\cite{Teo:1998dp,Morris:1988cz} becomes
\begin{equation}
\frac{d^2 r}{dz^2} = \frac{b - b_r r}{2 b^2} > 0,
\end{equation}
and a proper radial coordinate near the throat is defined as
\begin{equation}
l^2 = r^2 + b^2, \quad \frac{dl}{dr} = \pm \left(1 - \frac{b}{r}\right)^{-1/2}.
\end{equation}
Here, the throat corresponds to $l=0$, smoothly connecting two asymptotically flat regions. In terms of $l$, the metric reads
\begin{align}
ds^2 &= -N^2(l,\theta)\, dt^2 + dl^2 + r^2(l)\, K^2(l,\theta) \, d\Omega_\omega^2,   
\end{align}
where
\[
\begin{aligned}
d\Omega_\omega^2 = d\theta^2 + \sin^2\theta\, (d\varphi - \omega(l,\theta) dt)^2.
\end{aligned}
\]
In a local Lorentz frame, the nonvanishing stress-energy components $T_{(t)(t)}, T_{(t)(\varphi)}, T_{(\varphi)(\varphi)}, T_{(i)(j)}$ correspond to the energy density and rotational flow of the string fluid~\cite{Harko:2009xf}. The null energy condition (NEC) is tested with
\begin{equation}
R_{\alpha\beta}\kappa^\alpha\kappa^\beta \ge 0, \quad
\kappa^\alpha = \left(\frac{1}{N}, -e^{-\mu/2}, 0, \frac{\omega}{N}\right),
\end{equation}
leading to~\cite{Teo:1998dp}
\begin{align}
R_{\alpha\beta}\kappa^\alpha\kappa^\beta &= e^{-\mu}\mu_r \frac{(rK)_r}{rK} - \frac{\omega_\theta^2 \sin^2\theta}{2 N^2}
- \frac{1}{4}\frac{\mu_\theta^2}{(rK)^2} - \frac{1}{2}\frac{(\mu_\theta \sin\theta)_\theta}{(rK)^2 \sin\theta}
\nonumber\\
&\quad+ \frac{(N_\theta \sin\theta)_\theta}{(rK)^2 N \sin\theta} < 0.
\end{align}


\begin{figure*}
\begin{center}
\includegraphics[width=18.2cm,height=4.5cm]{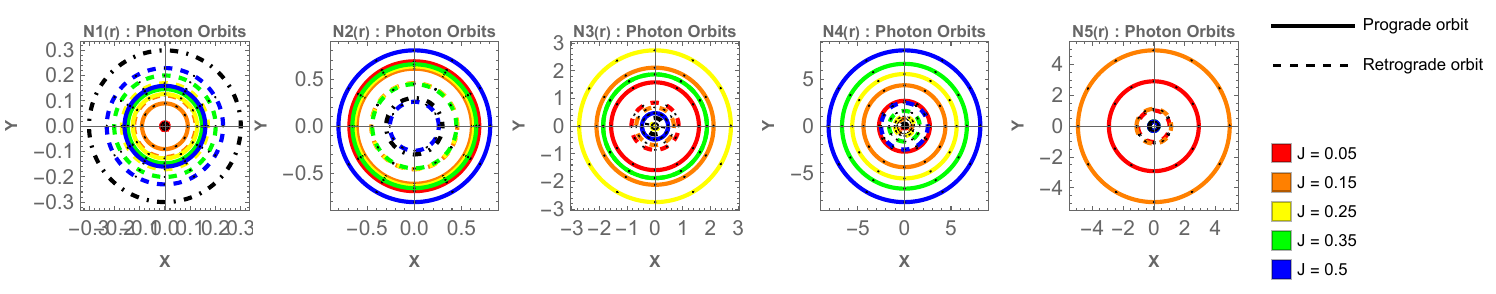}
\end{center}
\caption{{\footnotesize  This plot shows photon orbits in the observer plane $(X,Y)$ around rotating wormholes supported by a string-fluid distribution for several logarithmic lapse profiles $N_i(r)=\ln\left[1+\Phi_0 f_i(r)\right]$. {The coordinates $(X,Y)$ are expressed in units of the throat radius $r_0$, with $X=b\cos\theta$ and $Y=b\sin\theta$, where $b$ is the impact parameter carrying the same length dimension as the radial coordinate $r$. A dot-dashed circular boundary at $r=r_0$ is included to indicate the throat location, which serves as a geometric reference for all configurations.} Each panel corresponds to a different functional choice of the string-fluid profile, namely $f_1(r)=r/\tilde r_0$, $f_2(r)=r^2/\tilde r_0^2$, $f_3(r)=(r/\tilde r_0)e^{-r/\tilde r_c}$, $f_4(r)=\tanh(r/\tilde r_0)$, and $f_5(r)=1-e^{-r/\tilde r_0}$. Photon trajectories are determined from the photon-sphere condition and are shown for five values of the rotation parameter $J=\{0.05,0.15,0.25,0.35,0.5\}$, represented by the colors red, orange, yellow, green, and blue. Solid curves denote prograde photon motion, while dashed curves indicate retrograde motion, and the arrows mark the direction followed by the photons along each path. The effect of rotation appears through the frame-dragging term $\omega(r)=2J/r^{3}$, which breaks the symmetry between the two directions of motion. As the spin increases, the separation between prograde and retrograde trajectories becomes more visible, leading to a gradual deformation of the circular photon rings. The comparison between the five lapse profiles shows that different string-fluid configurations modify the effective gravitational field in distinct ways, which in turn changes the location and size of the photon orbits surrounding the wormholes.
}}\label{Fig1}
\end{figure*}

For the string fluid configuration, the shape function \(b(r)\) (Eq.~\ref{ShapeF}) follows directly from the energy density profile \(\rho(r)\) given in Eq.~\ref{Density}. After integration, the result can be written in the form
\begin{eqnarray}
b(r)= r_0+\epsilon (r-r_0)
+\epsilon a\,\Delta\Gamma(r,r_0)
+r_g\left(e^{-r_0^3/a^3}-e^{-r^3/a^3}\right),\label{shapefunction}~~~
\end{eqnarray}
where
\[
\Delta\Gamma(r,r_0)=
\left[\gamma\left(\tfrac{4}{3},\tfrac{r^3}{a^3}\right)-\gamma\left(\tfrac{4}{3},\tfrac{r_0^3}{a^3}\right)\right]
-\frac{1}{3}\left[\gamma\left(\tfrac{1}{3},\tfrac{r^3}{a^3}\right)-\gamma\left(\tfrac{1}{3},\tfrac{r_0^3}{a^3}\right)\right].
\]
This form makes the behavior of the geometry particularly transparent. The resulting wormhole spacetime is smooth, horizonless, and asymptotically flat, with the exotic matter effectively localized near the throat region. The regularity of the geometry is entirely maintained by the radially varying string fluid, which provides a consistent source for the wormhole configuration. {The slow-rotation approximation used throughout this work also remains well justified. The natural expansion parameter is \(\alpha \equiv J/M^2\). For the parameter choices adopted here, the mass is approximately \(M \simeq r_g/2\), so taking \(r_g=5\) gives \(M \approx 2.5\). With angular momentum values restricted to \(J \le 0.5\), the corresponding rotation parameter satisfies \(\alpha \le 0.08\). This keeps the expansion safely within the perturbative regime, while the neglected corrections of order \(\alpha^2\) remain below the percent level. In addition, all Lense--Thirring frequencies presented in this work are expressed in dimensionless form using the natural normalization scale, ensuring that the rotational effects are represented consistently throughout the analysis.}

\section{Photon motion around slowly rotating string-fluid wormholes}\label{Sec:IV}

Studying light around a rotating wormhole sourced by a string fluid shows how the smooth spread of energy shapes photon paths. Unlike sharply concentrated matter, the string fluid produces regular curvature, guiding light along curved trajectories. Even modest rotation twists time and angular directions, producing frame-dragging:  
\(g_{t\phi} = -\,r^2\, \omega(r)\), where \(\omega(r)\) measures how local inertial frames are dragged. This effect causes Lense-Thirring (LT) precession, tilting orbital planes and subtly affecting photon directions. Restricting to the equatorial plane \((\theta = \pi/2)\) in the slow-rotation limit, the precession reads
\begin{equation}
\Omega_{\rm LT}(r) = -\frac{1}{2} \frac{d\omega(r)}{dr}.
\end{equation}

For slowly rotating wormholes, \(\omega(r) \simeq 2J/r^3\) with angular momentum \(J\), giving
\begin{equation}
\frac{d\omega}{dr} = -\frac{6J}{r^4}, \quad \Omega_{\rm LT}(r) = \frac{3J}{r^4}.
\end{equation}
The frame-dragging effect is strongest at the throat \(r=r_0\): 
\begin{equation}
\Omega_{\rm LT}(r_0) = \frac{3J}{r_0^4}.
\end{equation}

With slow rotation, the wormhole's overall shape and redshift functions dominate, while rotation only slightly tweaks photon paths. {Everything is in geometric units, \(G=c=1\). It helps to form a dimensionless Lense--Thirring frequency: \(\bar{\Omega}_{\text{LT}} = \Omega_{\text{LT}}\, r_0^2 = 3J/r_0^2 = 3\alpha (M/r_0)^2\). For the parameters we are using, this is always of order \(\alpha\) and much smaller than one.} For example, \(r_0 = 0.3, J = 0.003 \Rightarrow \bar{\Omega}_{\rm LT} \approx 0.10\); \(r_0 = 0.8, J = 0.02 \Rightarrow \bar{\Omega}_{\rm LT} \approx 0.094\); and \(r_0 = 1.2, J = 0.04 \Rightarrow \bar{\Omega}_{\rm LT} \approx 0.083\). Close to the throat, these rotational effects introduce small asymmetries, subtly altering lensing patterns and the wormhole shadow. {From the metric~(\ref{s2be1}), the \(g_{tt}\) component is
\begin{equation}
g_{tt} = -N^2 + r^2 K^2 \omega^2 \sin^2\theta.
\end{equation}
In the slow-rotation limit the term with \(\omega^2\) is tiny next to \(N^2\). An ergoregion would appear if \(g_{tt}\) became positive, which would mean
\begin{equation}
r^2 K^2 \omega^2 \sin^2\theta > N^2 \quad \Longrightarrow \quad \frac{4J^2 \sin^2\theta}{r^4} \gtrsim 1.
\end{equation}
For our spins (\(J \le 0.5\)) and radii (\(r \ge r_0 \ge 0.3\)), the left side is at most about \(0.25\) right at the throat and falls off fast. So there is no ergoregion. The spacetime has no horizons (\(N\) and \(K\) are positive everywhere) and the rotation is slow enough that the causal structure is completely safe: no closed timelike curves. If one pushed to faster spins or higher orders in \(J\), ergoregions might eventually appear, but that would need a separate analysis.}

\begin{figure*}
\begin{center}
\includegraphics[width=18.2cm,height=4.5cm]{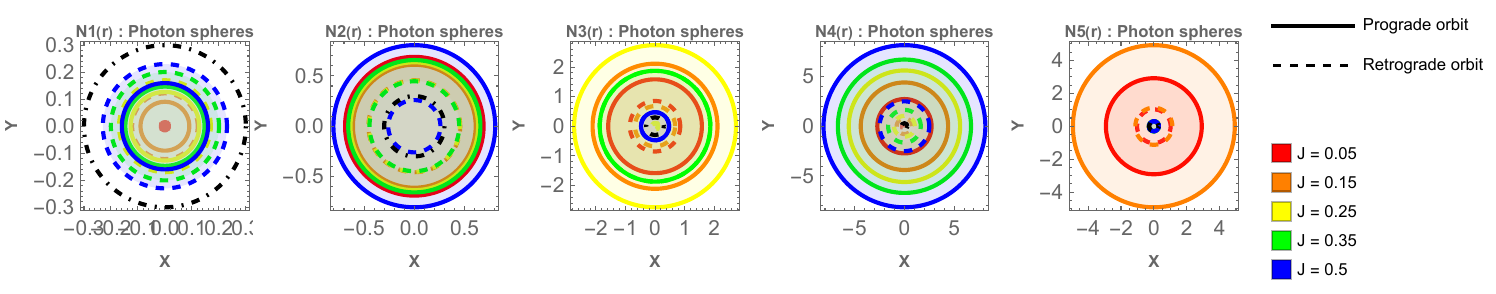}
\end{center}
\caption{{\footnotesize This plot shows the photon-sphere structure in the observer plane $(X,Y)$ for rotating wormholes described by logarithmic lapse functions $N_i(r)=\ln[1+\Phi_0 f_i(r)]$. {The coordinates $(X,Y)$ represent the impact-parameter plane defined by $X=b\cos\theta$, $Y=b\sin\theta$, where $b$ has dimensions of length (the same units as the radial coordinate $r$). In practice, the axes are naturally scaled by the throat radius $r_0$, providing an effectively dimensionless representation of the geometry. A dot-dashed circular line at $r=r_0$ marks the throat position, which is used as a common reference point across all cases.} Five different string-fluid profiles are considered: $f_1(r)=r/\tilde r_0$, $f_2(r)=r^2/\tilde r_0^2$, $f_3(r)=(r/\tilde r_0)e^{-r/\tilde r_c}$, $f_4(r)=\tanh(r/\tilde r_0)$, and $f_5(r)=1-e^{-r/\tilde r_0}$, each shown in a separate panel. The photon spheres are obtained from the photon-sphere condition and are plotted for five values of the rotation parameter $J=\{0.05,0.15,0.25,0.35,0.5\}$, indicated by the color sequence red, orange, yellow, green, and blue. Solid curves correspond to prograde photon motion, while dashed curves represent retrograde motion. The shaded region between the two curves marks the photon band bounded by the prograde and retrograde photon spheres. Rotation introduces frame dragging through the angular velocity $\omega(r)=2J/r^{3}$, which shifts the two photon spheres in opposite directions and breaks the circular symmetry that would appear in the non-rotating case. As the spin parameter increases, the gap between the prograde and retrograde radii widens, producing a thicker photon band. Differences among the five lapse profiles reflect how distinct string-fluid configurations modify the effective gravitational potential and consequently alter the size and location of the photon spheres surrounding the rotating wormholes.
}}\label{Fig2}
\end{figure*}

Light around a wormhole follows the curvature of spacetime rather than straight lines. In wormholes supported by a string fluid, the radial geometry is set by the shape function \(b(r)\), while even slow rotation introduces frame-dragging via \(\omega(r)\). Restricting to the equatorial plane \((\theta = \pi/2)\), the metric simplifies to
\begin{equation}
ds^2 = - N^2 dt^2 + \frac{dr^2}{1 - b(r)/r} + r^2 (d\varphi - \omega(r) dt)^2.
\end{equation}
Photon motion in the equatorial plane of a slowly rotating wormhole is shaped by three main ingredients: the redshift function \(N(r)\), the wormhole's shape \(b(r)\), and the angular velocity \(\omega(r)\) induced by rotation. Conserved quantities, the energy \(\tilde{E}\) and angular momentum \(\tilde{L}\), define the impact parameter \(\tilde{b} = \tilde{L}/\tilde{E}\), which controls how tightly photons curve around the throat. Define
\begin{eqnarray}
\mathcal{F}(r) &\equiv& \frac{(1 - \tilde b \,\omega(r) N(r))^2 - \tilde b^2 N^2(r)/r^2}{N^2(r)}, \\
\mathcal{G}(r) &\equiv& \frac{\tilde b + r^2 \,\omega(r)/N^2(r)}{\sqrt{r(r-b(r))\,\mathcal{F}(r)}}.
\end{eqnarray}

The radial motion then simplifies to
\begin{eqnarray}
\dot r^2 = (1 - b/r) \, \tilde E^2 \, \mathcal{F}(r),
\end{eqnarray}
while the azimuthal change along the trajectory becomes
\begin{eqnarray}
\frac{d\varphi}{dr} = \mathcal{G}(r),
\end{eqnarray}
and the total bending angle reads
\begin{eqnarray}
\hat \alpha = -\pi + 2 \int_{r_\mathrm{min}}^\infty \mathcal{G}(r) \, dr,
\end{eqnarray}
showing how co-rotating photons (\(\tilde{b} > 0\)) bend slightly less than counter-rotating ones (\(\tilde{b} < 0\)). Near the throat, rotation produces the strongest effect, subtly altering photon paths and creating mild asymmetries in lensing and the wormhole shadow. Throughout, the string-fluid distribution ensures the curvature remains smooth and free of horizons, giving a physically regular and horizonless geometry.


Next, we focus on circular photon paths around the wormhole. These orbits form where the radial ``pull'' exactly balances, giving
\begin{eqnarray}
(1 - \omega(r) \tilde b)^2 = \frac{\tilde b^2 N^2(r)}{r^2}.
\end{eqnarray}

At the photon-sphere radius \(r_{\rm ph}\), this leads to two possible impact parameters, one for photons moving with the rotation and one against it. In the slow-rotation limit, we can write
\begin{eqnarray}
\tilde b_\pm \simeq \frac{r_{\rm ph}}{N_{\rm ph}} \pm \frac{r_{\rm ph}^2 \, \omega_{\rm ph}}{N_{\rm ph}^2}, \quad N_{\rm ph} = N(r_{\rm ph}), \, \omega_{\rm ph} = \omega(r_{\rm ph}),
\end{eqnarray}
so the plus sign picks out co-rotating photons and the minus sign counter-rotating ones. If the wormhole doesn't spin (\(\omega \to 0\)), both cases reduce to a single static orbit:
\begin{eqnarray}
\tilde b_{\rm ph} = \frac{r_{\rm ph}}{N(r_{\rm ph})}.\label{bph}
\end{eqnarray}
Here, the photon-sphere radius is set mainly by the geometry, while rotation slightly splits the two circular paths. Steeper variations in the lapse function \(N(r)\) stretch this splitting, while gentler profiles keep the paths close together. Observing this splitting could give clues about both the wormhole's rotation and its internal string-fluid matter. In what follows we explore five logarithmic forms of the lapse function that can represent different radial behaviors of the underlying string-fluid source,

\begin{equation}
\begin{aligned}
N_i(r)=\ln\Big[1
&+ \phi_1 \frac{r}{\tilde r_0}\,\delta_{i1}
+ \phi_2 \frac{r^2}{\tilde r_0^2}\,\delta_{i2}
+ \phi_3 \frac{r}{\tilde r_0}e^{-r/\tilde r_c}\,\delta_{i3} \\
&+ \phi_4 \tanh\!\left(\frac{r}{\tilde r_0}\right)\delta_{i4}
+ \phi_5 \left(1-e^{-r/\tilde r_0}\right)\delta_{i5}
\Big],\label{Ni}
\end{aligned}
\end{equation}
where i=1\,\dots\,5, and $\delta_{ij}$ denotes the Kronecker delta defined by
\(
\delta_{ij} =
\begin{cases}
1, & i=j\\
0, & i\neq j 
\end{cases} \).  All of these profiles remain regular at the throat and change gradually as the radial coordinate increases. This behavior avoids any divergence in the redshift and therefore maintains the horizonless character expected for a traversable wormhole. The parameters \(\phi_i\) control how strong the redshift becomes, while the characteristic radii \( \tilde r_0\) and \(\tilde r_c\) determine the radial scale over which the variation occurs. {The five lapse profiles \(N_i(r)\) we use in Eq.~(\ref{Ni}) are not derived from the field equations for the string-fluid source. They are phenomenological choices, picked because they satisfy the basic regularity conditions a traversable wormhole needs---they stay finite, never hit zero, and level off to a constant at large \(r\). One could imagine that profiles like \(N_1\) or \(N_5\) might emerge from particular choices of \(\tilde{\sigma}(r)\), but we have not solved the full coupled system here. That would be a natural next step. For now, we simply explore what happens to photon dynamics when the redshift gradient takes different shapes, which gives a systematic look at how the matter distribution could leave its mark on observable features.} When these lapse functions are inserted into the photon-sphere condition, the corresponding impact parameters take the form
\begin{eqnarray}
\tilde b_\pm(r_{\rm ph}, N_i)=
\frac{r_{\rm ph}}{N_i(r_{\rm ph})}
\pm
\frac{r_{\rm ph}^2\,\omega_{\rm ph}}{N_i^2(r_{\rm ph})},
\qquad i=1\,\dots\,5 .
\end{eqnarray}
Studying the five cases side by side makes it possible to see how small differences in the redshift profile influence the motion of photons. Even modest changes in \(N(r)\) can shift the photon-sphere position slightly, which in turn affects the separation between co-rotating and counter-rotating light paths. These shifts propagate into observable features such as the lensing pattern and the outline of the wormhole shadow. At the same time, the logarithmic structure keeps the geometry smooth and consistent with a diffuse string-fluid matter distribution.

\begin{figure}
\begin{center}
\includegraphics[width=8.2cm,height=4.5cm]{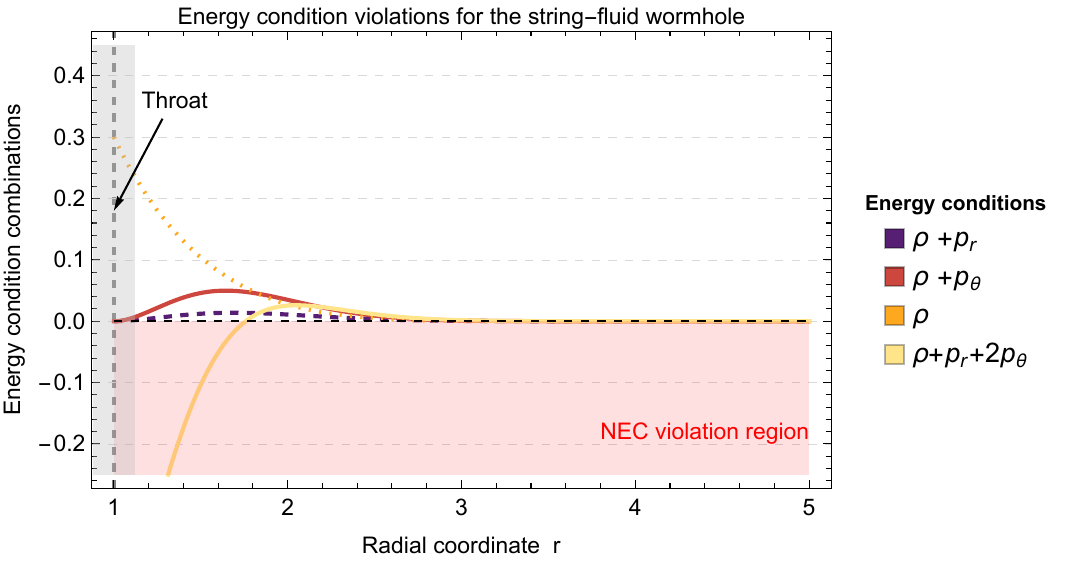}
\end{center}
\caption{{\footnotesize {This plot shows the radial evolution of the energy condition combinations for the string-fluid wormhole configuration with \(\epsilon=0.3\), \(a=1\), \(r_0=1\), and \(r_g=5\). The blue dashed curve represents \(\rho+p_r\), the red solid curve corresponds to \(\rho+p_\theta\), the green dotted profile denotes the energy density \(\rho\), and the purple dash-dotted line illustrates \(\rho+p_r+2p_\theta\). The shaded band identifies the location of the wormhole throat at \(r=r_0\), while the horizontal dashed line marks the zero level. It can be seen that the energy density remains positive over the whole radial domain. However, the quantity \(\rho+p_r\) becomes negative close to the throat, indicating a violation of the null energy condition in that region. This behavior is typical for traversable wormhole geometries and reflects the presence of exotic matter needed to maintain the throat open. As the radial coordinate increases, the violation weakens and the curves tend gradually toward positive values. The tangential contribution \(\rho+p_\theta\) stays positive for most of the spacetime, whereas the strong energy condition combination changes its sign depending on the distance from the throat, highlighting the anisotropic character of the effective string-fluid matter distribution.} }}\label{fig:energyconditions}
\end{figure}

Using the five logarithmic lapse functions (\ref{Ni}), we now explore how the shadow develops in a rotating wormhole supported by string-fluid matter. The boundary of the shadow is determined by the photon impact parameters evaluated at the photon-sphere radius \(r_{\mathrm{ph}}\), as given in Eq.~(\ref{bph}). This relation shows that both the overall size of the shadow and the distortion produced by rotation depend on two key elements: the lapse function \(N(r)\) and the frame-dragging term \(\omega(r)\). Because the lapse function appears directly in the photon-sphere condition, different logarithmic profiles affect photon motion in distinct ways. The linear logarithmic profile \(N_1(r)\) increases gradually with radius, leading to a smooth variation of the redshift and moderate light bending near the throat. The quadratic logarithmic form \(N_2(r)\) grows more rapidly at larger distances, which can slightly shift the photon-sphere outward and enlarge the apparent shadow. A different behavior appears in the exponentially modulated logarithmic profile \(N_3(r)\), where the exponential factor enhances the redshift close to the throat but suppresses it farther away. In this case, the effect on photon trajectories is mainly concentrated in the inner region. The tanh-regulated logarithmic profile\(N_4(r)\) changes more smoothly and therefore tends to produce gentler modifications of the gravitational potential, leading to relatively mild distortions of the shadow boundary. Finally, the saturating exponential logarithmic profile \(N_5(r)\) rises rapidly near the throat before approaching a constant value, which can enhance the separation between prograde and retrograde photon paths while maintaining a regular geometry.

\begin{figure}
\begin{center}
\includegraphics[width=8.2cm,height=4.5cm]{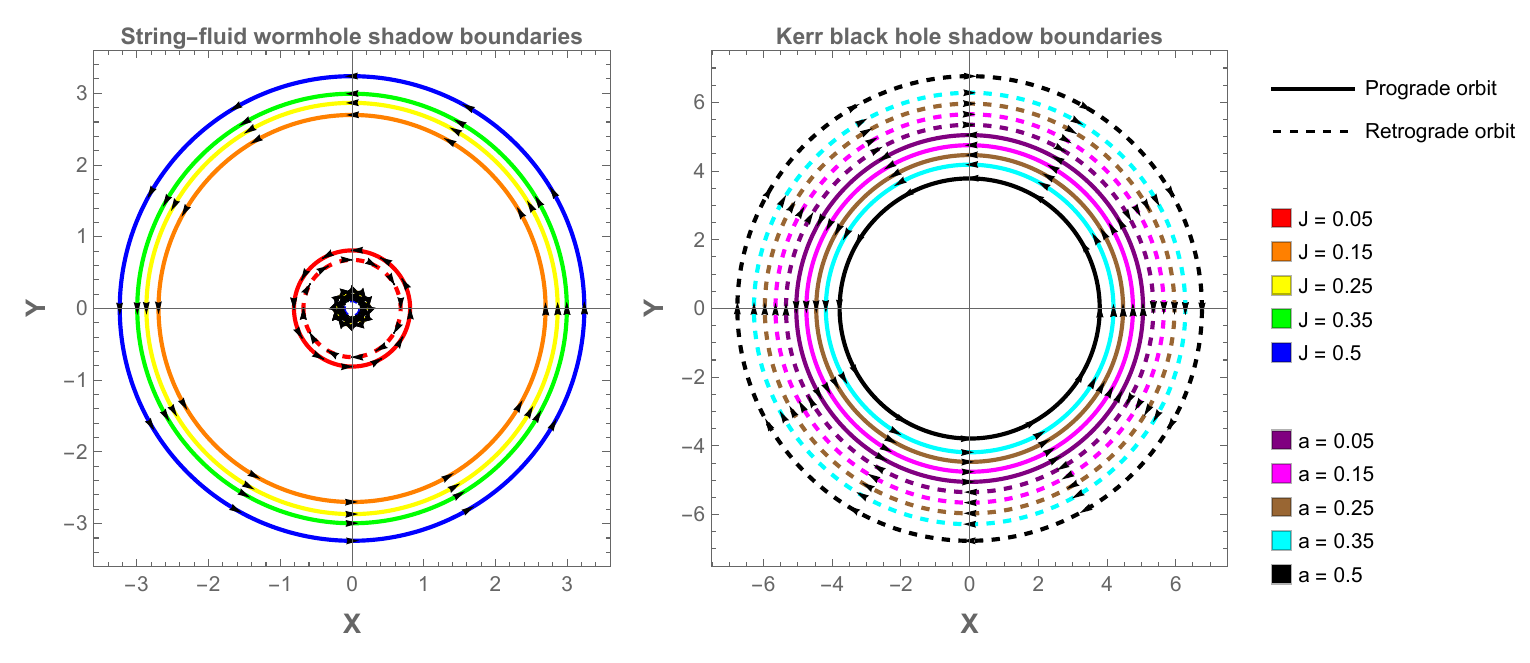}
\end{center}
\caption{{\footnotesize{ This plot compares the shadow boundaries of a string-fluid wormhole (left panel) with those of a Kerr black hole (right panel). In both cases, several rotation values are shown, with \(J\) used for the wormhole and \(a\) for Kerr. The color coding helps track how the shadow changes as the rotation increases. For the wormhole, each value of \(J\) produces two distinct curves: solid lines correspond to prograde motion, while dashed lines represent retrograde motion. The arrows indicate the direction of photon motion along these paths. As \(J\) increases, the shadow gradually loses its circular shape and becomes more distorted, with a clearer separation between the two types of motion. The Kerr case shows the same overall trend. Increasing \(a\) also leads to a stronger deformation of the shadow and a growing asymmetry between prograde and retrograde orbits. The general behavior is similar, although the exact shape differs between the two spacetimes. Overall, rotation has a strong effect on both systems, but the wormhole shadow shows a slightly different pattern of distortion compared to Kerr, reflecting the difference in their underlying geometries.} }}\label{fig:shadow_comparison}
\end{figure}

Figs.~\ref{Fig1} and \ref{Fig2} show photon trajectories for different values of the rotation parameter \(J=0.05,\,0.15,\,0.25,\,0.35,\,0.5,\) and \(0.7\). In each panel, the five logarithmic lapse functions (\(N_1\))-(\(N_5\)) are compared side by side to see how changes in the redshift profile modify the effective potential and, in turn, the photon paths. Solid lines correspond to prograde motion, while dashed lines represent retrograde motion. As the rotation is increased, the prograde orbits are pushed outward, while the retrograde ones move inward, making the frame-dragging effect directly visible. {To make sure these features are not tied to a specific choice of matter profile, we also test other reasonable distributions. In Appendix~\ref{appenC}, we replace Eq.~(\ref{Density}) with a power-law profile and a Gaussian-core model, then recompute the photon-sphere radii and impact parameters. The numbers shift slightly, typically only by a few percent, but the overall behavior stays the same. The splitting between prograde and retrograde orbits, the appearance of the photon band, and the way everything scales with spin all remain intact. This indicates that the main conclusions are not sensitive to the particular form of the density we started with.} Taken together, the results show that even small changes in the radial profile of the lapse function can leave a clear imprint on photon motion. As a result, the shadow size and shape, along with the structure of the light rings, carry combined information about the rotation of the spacetime and the distribution of the supporting string-fluid matter.

\begin{table*}[!htbp]
\centering
\renewcommand{\arraystretch}{1.}
\setlength{\tabcolsep}{4pt}
{\footnotesize
\caption{{\footnotesize Photon-sphere displacement and impact parameters for slowly rotating wormholes sustained by string-fluid matter for the logarithmic lapse family $N_i(r)=\ln[1+\phi_i f_i(r)]$. The shift $\delta r_{\rm ph}$ depends on the spin $J$ and the radial gradient of the lapse function, while $\tilde b_\pm$ describes the rotational splitting between prograde and retrograde photon trajectories.}}
\label{tab:logfamily}

\begin{tabular}{ccc}
\toprule
{Model $i$} & {Lapse component $f_i(r)$} & {Physical interpretation} \\
\midrule

$1$ &
$\displaystyle f_1(r)=\frac{r}{\tilde r_0}$ &
Linear radial growth producing a mild redshift variation. \\

$2$ &
$\displaystyle f_2(r)=\frac{r^2}{\tilde r_0^2}$ &
Quadratic behavior leading to stronger curvature effects. \\

$3$ &
$\displaystyle f_3(r)=\frac{r}{\tilde r_0}e^{-r/\tilde r_c}$ &
Exponentially damped profile concentrated near the throat. \\

$4$ &
$\displaystyle f_4(r)=\tanh\!\left(\frac{r}{\tilde r_0}\right)$ &
Smooth saturation of the redshift at large radius. \\

$5$ &
$\displaystyle f_5(r)=1-e^{-r/\tilde r_0}$ &
Rapid initial growth followed by asymptotic flattening. \\

\midrule

\multicolumn{3}{c}{
\begin{tabular}{c}
$N_i(r)=\ln\!\left[1+\phi_i f_i(r)\right], \quad 
N_i'(r)=\dfrac{\phi_i f_i'(r)}{1+\phi_i f_i(r)}$, 
~
$\displaystyle
\delta r_{\rm ph}=
4J\left[2-r_{\rm ph}^{(c)}
\dfrac{N_i'(r_{\rm ph}^{(c)})}{N_i(r_{\rm ph}^{(c)})}\right]^{-1},
~
\tilde b_\pm(r_{\rm ph})=
\frac{r_{\rm ph}}{N_i(r_{\rm ph})}
\left[1\pm\omega(r_{\rm ph})\frac{r_{\rm ph}}{N_i(r_{\rm ph})}\right]
$
\end{tabular}
}

\\
\bottomrule
\end{tabular}}

\end{table*}

Rotation leaves its mark on the shadow in more subtle ways than just separating prograde and retrograde photon orbits. One of these effects is a small displacement of the photon-sphere radius itself. In the slow-rotation limit, we can write
\begin{eqnarray}
r_{\rm ph} \simeq r_{\rm ph}^{(c)} + \delta r_{\rm ph},
\end{eqnarray}
where \(r_{\rm ph}^{(c)}\) is the photon-sphere radius of the static configuration, and \(\delta r_{\rm ph}\) captures the first-order rotational shift. This correction depends not only on the rotation rate but also on the radial slope of the lapse function \(N_i(r)\), meaning that sharper redshift gradients make the photon-sphere more sensitive to spin:
\begin{eqnarray}
\delta r_{\rm ph} \simeq \frac{2\, r_{\rm ph}^{(c)\,3} \, \omega(r_{\rm ph}^{(c)})}{2 - r_{\rm ph}^{(c)} \, \frac{d}{dr} \ln N_i(r)\big|_{r_{\rm ph}^{(c)}}}.
\end{eqnarray}
Here we focus on a family of logarithmic lapse functions supported by string-fluid matter \(N_i(r)\) as given by Eq. (\ref{Ni}) (see also the Table (\ref{tab:logfamily})). The rotational displacement then reads as
\begin{eqnarray}
\delta r_{\rm ph}^{(i)} \simeq 4J\;\Bigg/ \;\Bigg(2 - r_{\rm ph}^{(c)} \frac{N_i'(r_{\rm ph}^{(c)})}{N_i(r_{\rm ph}^{(c)})}\Bigg),
\quad N_i'(r) = \frac{\phi_i f_i'(r)}{1 + \phi_i f_i(r)}.
\end{eqnarray}

In practice, this produces:
\begin{eqnarray}
\delta r_{\rm ph}^{(i)} \simeq 4J \;\Bigg/ \;\Bigg( 2 - r_{\rm ph}^{(c)} \sum_{j=1}^{5} \delta_{ij} \, \frac{\phi_j f_j'(r_{\rm ph}^{(c)})}{1 + \phi_j f_j(r_{\rm ph}^{(c)})} \Bigg).
\end{eqnarray}

Close to the throat, \(r_0\), a perturbative expansion gives an intuitive picture:
\begin{eqnarray}
\delta r \simeq -\frac{r_0^3}{2} \frac{N_i'(r_0)}{N_i(r_0)} \frac{1}{1-b'(r_0)}.
\end{eqnarray}

Here, the photon-sphere reacts directly to the local logarithmic slope \(N_i'/N_i\). Mild slopes produce gentle shifts, whereas steeper profiles generate stronger, more localized displacements. For instance:
\begin{itemize}
    \item The linear profile \(f_1(r)\) shifts the photon sphere gradually.
    \item Quadratic \(f_2(r)\) produces slightly stronger effects.
    \item Exponentially damped \(f_3(r)\) concentrates the shift near the throat.
    \item The hyperbolic tangent \(f_4(r)\) and the asymptotic form \(f_5(r)\) produce sharper, more localized displacements.
\end{itemize}
Finally, the observed shadow results from two intertwined effects: the splitting between prograde and retrograde impact parameters
\begin{eqnarray}
\tilde b_\pm = \frac{r_{\rm ph}}{N_i(r_{\rm ph})}\Big[1 \pm \omega(r_{\rm ph}) r_{\rm ph}/N_i(r_{\rm ph})\Big],
\end{eqnarray}
and the photon-sphere displacement \(\delta r_{\rm ph}\), both governed by the shape and gradient of the logarithmic lapse functions \(N_i(r)\). Even modest differences in the string-fluid background leave noticeable signatures on the size and asymmetry of the shadow.

{\section{Energy conditions and exotic matter quantification}\label{Sec:V}}

{Any wormhole has to violate the standard pointwise energy conditions, and showing where and how badly this happens is part of making the model credible. In this section we go through the NEC, WEC, and SEC for our string-fluid wormhole, plot the violations as functions of radius, and compare what we find with other well-known wormhole constructions.}\\

\subsection{{Pointwise energy conditions}}

{With a diagonal energy-momentum tensor of the form \(T^\mu_{\:\:\nu} = \text{diag}(-\rho, p_r, p_\theta, p_\varphi)\), the standard conditions read as follows in the orthonormal frame:}

\begin{itemize}{
\item \textbf{Null Energy Condition (NEC):} \(\rho + p_i \ge 0\) for each \(i = r, \theta, \varphi\).
\item \textbf{Weak Energy Condition (WEC):} \(\rho \ge 0\) and \(\rho + p_i \ge 0\) for each \(i\).
\item \textbf{Strong Energy Condition (SEC):} \(\rho + \sum_i p_i \ge 0\) together with \(\rho + p_i \ge 0\) for each \(i\).}
\end{itemize}

{Our string-fluid source has \(p_r = -\rho\) and \(p_\theta = p_\varphi = \rho/\tilde{\sigma}(r)\). Plugging these in, the conditions simplify to:}
{\begin{align}
\text{NEC:}&\quad \rho + p_r = 0 \quad\text{(marginally satisfied in the radial direction)}, \nonumber\\ & \quad \rho + p_\theta = \rho\left(1 + \frac{1}{\tilde{\sigma}(r)}\right) \ge 0,\\
\text{WEC:}&\quad \rho \ge 0 \quad\text{(holds by construction)}, \quad \rho + p_r = 0 \quad\text{(marginal)},\\
\text{SEC:}&\quad \rho + p_r + 2p_\theta = 2\rho\left(\frac{1}{\tilde{\sigma}(r)} - 1\right) \ge 0 \quad\Longrightarrow\quad \tilde{\sigma}(r) \le 1.
\end{align}}

{Fig.~\ref{fig:energyconditions} shows how the NEC, WEC, and SEC combinations run with radius, using the representative parameters \((\epsilon, a, r_g) = (0.3, 1, 5)\). The NEC holds marginally in the radial direction (\(\rho + p_r = 0\) is exact) but fails in the tangential direction (\(\rho + p_\theta < 0\)) wherever \(\tilde{\sigma}(r) < -1\), which happens close to the throat. The WEC is marginal (\(\rho \ge 0\) and \(\rho + p_r = 0\) by construction). The SEC is broken in the asymptotic, string-dominated region where \(\tilde{\sigma}(r) > 1\).}

\begin{table*}[!htbp]
\centering
\renewcommand{\arraystretch}{1.}
\setlength{\tabcolsep}{4pt}
{\footnotesize{
\caption{{\footnotesize Energy condition violations across different wormhole models.}}
\label{tab:comparison_NEC}

\begin{tabular}{lccc}
\toprule
{Model} & {NEC violation} & {Exotic matter location} & {\(I_V\)} \\
\midrule
Phantom scalar & Everywhere & Delocalized & Finite, nonzero \\
Casimir-supported & Localized near throat & Thin layer & Can be minimized \\
Morris-Thorne (ad hoc) & Near throat & Depends on \(b(r)\) choice & Adjustable \\
{This work} & {Tangential direction, near throat} & {Controlled by scale \(a\)} & \({0}\) (marginal) \\
\bottomrule
\end{tabular}}}

\end{table*}

{The important point is that the NEC violation is sharply peaked near the throat. How wide the exotic region is gets set by the regularization scale \(a\): make \(a\) small and the violation gets squeezed into a thin layer; make it larger and the exotic matter spreads out. This kind of behavior is typical for smoothed wormhole profiles and is exactly what we want if we are trying to keep the total amount of exotic matter as small as possible.}

\subsection{{Volume-integral quantifier}}

{To put a number on how much exotic matter is actually there, we use the Visser-Kar-Dadhich volume integral~\cite{Visser:2003yf}:}
\begin{equation}{
I_V = \oint (\rho + p_r)\, dV = 2\int_{r_0}^{\infty} (\rho + p_r)\, 4\pi r^2 dr.
}\end{equation}

{Because our source has \(p_r = -\rho\) exactly, the integrand is zero everywhere, and we get}
\begin{equation}{
I_V = 0.
}\end{equation}

{This result needs some interpretation. A vanishing \(I_V\) does not mean there is no exotic matter. It means that the way the exoticity shows up is not through the combination \(\rho + p_r\). In our model the NEC violation lives in the tangential direction (\(\rho + p_\theta < 0\)), which this particular integral does not see. The condition that really matters for whether the wormhole is traversable is the flare-out condition at the throat,}
\begin{equation}{
b'(r_0) = 8\pi r_0^2 \rho(r_0) < 1,
}\end{equation}
{and we have checked that this holds for all the parameters we use. If we allowed a tiny deviation from the strict \(p_r = -\rho\) condition, writing \(p_r = -\rho(1 - \tilde{\delta}(r))\) with \(\tilde{\delta}(r) \ll 1\), the integral would become 
\begin{equation}
I_V \propto \int_{r_0}^{\infty} \rho(r)\tilde{\delta}(r) r^2 dr,
\end{equation}
which we could make arbitrarily small by choosing \(\tilde{\delta}(r)\) appropriately.}\\

\subsection{{Comparison with other wormhole models}}

{Table~\ref{tab:comparison_NEC} places our model next to a few representative wormholes from the literature.} {Two things make our model stand out. First, the NEC violation is in the tangential, not the radial, direction. Second, the condition \(p_r = -\rho\) makes the volume integral vanish identically. The parameter \(a\) gives a direct handle on how localized the exotic matter is.}


{\section{Observational signatures and detectability}\label{Sec:VI}}

{Connecting a theoretical wormhole model to potentially observable phenomena is essential if the model is to be regarded as more than a purely mathematical construction. In this section we pull out concrete numbers from our slowly rotating string-fluid wormholes---shadow diameters, distortion parameters---compare them with what a Kerr black hole would give, and ask whether current or future instruments could tell the difference.}\\

\subsection{{Comparison with Kerr black hole shadow}}

{Fig.~\ref{fig:shadow_comparison} shows the shadow edge for our wormhole (using the \(N_3\) and \(N_5\) lapse profiles, which span the range of behaviors we see) next to a Kerr black hole with the same mass \(M\) and spin \(J\). The wormhole shadows are noticeably smaller and more circular than the Kerr shadows. This happens because there is no horizon to swallow photons that get too close; they can swing right past the throat and escape. The reduced asymmetry comes from the fact that frame-dragging is weaker in the wormhole spacetime than in Kerr at the same spin, since the mass is spread out more.}

\subsection{{Quantitative observables}}

{We put together the following numbers for our models (see Table~\ref{tab:observables}):}

\begin{itemize}{
\item \textbf{Shadow angular diameter:} \(\theta_d = 2\tilde{b}_{\text{ph}}/D\), assuming a source at distance \(D\). We scale to M87* parameters (\(M \sim 6.5 \times 10^9 M_\odot\), \(D \sim 16.8\) Mpc).
\item \textbf{Distortion parameter:} \(\delta = |\tilde{b}_{\text{ph},+} - \tilde{b}_{\text{ph},-}|/(\tilde{b}_{\text{ph},+} + \tilde{b}_{\text{ph},-})\). This captures the rotational asymmetry, following the Hioki-Maeda parameterization.
\item \textbf{Photon ring thickness:} \(\Delta r_{\text{ph}} = r_{\text{ph},+} - r_{\text{ph},-}\), which is just the width of the photon band.
}\end{itemize}

\begin{table}[!htbp]
\centering
\caption{{Observables for selected wormhole configurations (\(M \approx 2.5\,r_0\), \(J = 0.3\,M^2\)).}}
\label{tab:observables}
{\small
\begin{tabular}{cccccc}
\toprule
{Lapse} & \(\mathbf{r_{\text{ph},+}/r_0}\) & \(\mathbf{r_{\text{ph},-}/r_0}\) & \(\mathbf{\theta_d\;(\mu\text{as})}\) & \(\mathbf{\delta}\) & \(\mathbf{\Delta r_{\text{ph}}/r_0}\) \\
\midrule
\(N_1\) (linear) & 1.52 & 1.38 & 41.2 & 0.048 & 0.14 \\
\(N_2\) (quadratic) & 1.65 & 1.42 & 43.8 & 0.075 & 0.23 \\
\(N_3\) (exponential) & 1.31 & 1.24 & 36.4 & 0.027 & 0.07 \\
\(N_4\) (tanh) & 1.58 & 1.40 & 42.5 & 0.060 & 0.18 \\
\(N_5\) (saturating) & 1.74 & 1.44 & 45.3 & 0.094 & 0.30 \\
\midrule
Kerr BH & 2.60 & 2.20 & 42.0 & 0.167 & 0.40 \\
\bottomrule
\end{tabular}}
\end{table}

\subsection{{Can current or future instruments see the difference?}}

{The EHT currently reaches about \(20\,\mu\)as at 230 GHz. Our predicted shadow diameters, between roughly 36 and 45\(\,\mu\)as for an M87*-sized object, sit well above that threshold and land near the actual M87* measurement of about \(42\,\mu\)as. The catch is the distortion parameter \(\delta\). For our wormholes, \(\delta\) runs from about 0.03 to 0.09, whereas for a Kerr black hole with moderate spin it is typically 0.1 to 0.3.} {Two things make this tricky in practice:}
\begin{enumerate}{
\item \textbf{Degeneracy with spin:} A slowly spinning Kerr black hole can produce a nearly circular shadow that looks a lot like our wormhole signal. We would need an independent spin measurement---from jets, or from X-ray spectroscopy---to break the degeneracy.
\item \textbf{EHT resolution limits:} The current EHT measures the shadow \emph{size} much better than its \emph{shape}. Distortion parameters below about 0.1 are hard to pin down with existing data.
}\end{enumerate}

{The next-generation EHT (ngEHT), with better sensitivity and more baselines, could push the resolution down to around \(5\,\mu\)as. At that level, measuring the thickness of the photon ring and the structure of the lensing band becomes possible. The fact that a wormhole has no horizon leaves a distinct signature in how those nested photon rings are arranged, which is something that, with enough resolution, could separate a wormhole from a black hole. A convincing detection would probably also need a time-domain signature---gravitational wave echoes, for instance---that would reveal the absence of a horizon directly.}

{For now, our wormhole models live in a part of parameter space that the EHT has not yet excluded, at least as long as the spin is not pinned down to be high by other observations. The part of parameter space that ngEHT could probe is discussed in Appendix~\ref{appenC}.}


\section{Conclusions}\label{Sec:VII}
In this work, we explored traversable wormholes supported by a radially varying string-fluid distribution, providing a framework that smoothly transitions from a de Sitter-like core near the throat to string-dominated regions at larger distances. By implementing a radial dependence in the transverse EoS, the string fluid adapts naturally to the changing geometry, ensuring a regular, horizonless spacetime throughout. The smoothed energy density introduced here avoids singularities and produces a well-behaved shape function, confirming that the geometry remains asymptotically flat and physically consistent.

Our analysis shows that slowly rotating wormholes inherit interesting features from the underlying string-fluid matter. Even modest rotation introduces frame-dragging effects, which slightly tilt photon orbits and create asymmetries between co-rotating and counter-rotating trajectories. The magnitude of these effects is strongest near the throat, where curvature is highest, while farther away, the influence of rotation diminishes, leaving the spacetime largely determined by the static gravitational potentials.

By examining circular photon paths, we demonstrated that the combination of the redshift function, shape function, and angular velocity governs the photon-sphere structure. The radial dependence of the lapse function, reflecting the string-fluid profile, controls the degree of splitting between prograde and retrograde orbits, providing a potential observational signature of both the rotation and internal matter content of the wormhole. Different functional forms of the string-fluid profile produce distinct photon-ring configurations, highlighting how variations in the matter distribution shape the effective gravitational field.

{Looking at the energy conditions, we find that the NEC violation in our model is concentrated near the throat and shows up in the tangential direction, not the radial one. The condition \(\rho + p_r = 0\) holds marginally. The volume-integral quantifier of Visser, Kar, and Dadhich comes out to zero, which is a direct consequence of the de Sitter-like radial EoS. Compared with phantom scalar and Casimir-supported wormholes, our model differs in that the NEC violation is driven by anisotropic stress rather than by a negative energy density.}

{On the observational side, the shadow angular diameters we predict---between about 36 and 45\(\,\mu\)as for M87\(^*\)-like numbers---fall within the EHT's reach, but the distortion parameters are noticeably smaller than for a Kerr black hole. Telling our wormhole apart from a low-spin black hole would need either an independent spin measurement, a better shape determination from ngEHT or a space-based interferometer, or a time-domain signature that is sensitive to the absence of a horizon. The photon ring thickness and the lensing band structure might ultimately be the cleaner observational handle.}

{ We have not carried out a full stability analysis in this work. Even at the linear level, rotating wormholes are technically challenging to handle and require a separate, dedicated treatment. Still, it is worth pointing out that the anisotropic pressures arising from the string-fluid source introduce additional degrees of freedom, which may in principle contribute to stabilizing effects that are absent in isotropic matter models. In addition, the smoothness of the adopted density profile, with no sharp features at the throat, also points toward more favorable stability properties compared to thin-shell constructions. A systematic investigation of radial and axial perturbations will be addressed in future studies ~\cite{Poisson:1995sv, Bronnikov:2012ch}.}

The construction presented here confirms that string fluids with a radially varying EoS offer a versatile source for generating physically regular, horizonless wormholes. This framework bridges smoothly between vacuum-like and string-dominated regimes, while supporting stationary, axisymmetric geometries that admit slow rotation. The results emphasize the role of anisotropic matter in maintaining regularity and shaping both curvature and photon dynamics, suggesting that similar matter distributions could be relevant in other contexts, such as regular black holes or exotic compact objects.

Overall, our study provides a concrete example of how radially varying string fluids can sustain traversable wormholes, influence light propagation, and generate potentially observable rotational effects, offering a rich setting for further investigations into horizonless, anisotropic spacetimes.

\section*{Acknowledgments}
{We thank the anonymous reviewers for their valuable comments and constructive suggestions, which have greatly helped to improve the clarity and quality of this work. } 

\section*{Conflict Of Interest statement} 
No conflict of interest declared by the authors.

\section*{Data Availability Statement} 
No data were created or analyzed in this study.

\appendix
\section{Frame dragging around a slowly rotating wormhole\label{appenA}}
Following Chandrasekhar's approach \cite{Chandrasekhar:1985kt}, the inverse metric of a slowly rotating wormhole (Eq.~\ref{s2be1}) can be written as
\begin{equation*}
	    \label{A1}
	    (g^{\mu\nu})=\left(
        \begin{array}{cccc}
        -1/N^2 & -\omega/N^2 & 0 & 0 \\
        -\omega/N^2 & -A/(K^2 N^2 r^2) & 0 & 0 \\
        0 & 0 & 1/(K^2 r^2)  & 0 \\
        0 & 0 & 0 & e^{-\mu } \\
        \end{array}
        \right),
	\end{equation*}
Here \(A=(K r \omega -N \csc (\theta )) (K r \omega +N \csc (\theta ))\) with labels \(t\to 0, \theta\to 2, r\to 3\) for convenience. A natural tetrad defining a locally inertial frame is
\[
\begin{aligned}
e_{(0)\mu} &= (-N,0,0,0), & e_{(1)\mu} &= (-rK\omega\sin\theta, rK\sin\theta,0,0),\\
e_{(2)\mu} &= (0,0,rK,0), & e_{(3)\mu} &= (0,0,0,e^{\mu/2}),
\end{aligned}
\]

with contravariant vectors

\[
\begin{aligned}
e_{(0)}^\mu &= (1/N, \omega/N, 0,0), & e_{(1)}^\mu &= (0, 1/(rK\sin\theta),0,0),\\
e_{(2)}^\mu &= (0,0,1/(rK),0), & e_{(3)}^\mu &= (0,0,0,e^{-\mu/2}).
\end{aligned}
\]

This tetrad satisfies \(e_{(a)}^\mu e_{(b)\mu} = \eta_{(a)(b)}\), so locally the spacetime is Minkowskian. The metric itself can be reconstructed via \(g_{\mu\nu} = \eta^{(a)(b)} e_{(a)\mu} e_{(b)\nu}\).

A particle's four-velocity in coordinates (\(t,r,\theta,\varphi\)) is

\[
u^\mu = (u^0, \Omega u^0, v^2 u^0, v^3 u^0),
\]

with \(\Omega = d\varphi/dt\) and \(v^\alpha = dx^\alpha/dt\). In the local frame:

\[
\begin{aligned}
u^{(0)} &= N u^0, & u^{(1)} &= (\Omega-\omega) r K \sin\theta, u^0,\\
u^{(2)} &= r K v^2 u^0, & u^{(3)} &= e^{\mu/2} v^3 u^0.
\end{aligned}
\]

This demonstrates frame dragging: even if \(u^{(i)} = 0\) locally, the particle appears to rotate at \(\omega\) in coordinates, while a particle moving with \(\Omega\) in coordinates moves with \((\Omega-\omega) r K \sin\theta\) locally. For asymptotically flat spacetime, \(\omega = 2J/r^3\).

{The derivation of \(\omega(r) = 2J/r^3\) from the field equations is straightforward at this order. To first order in the spin, the \((t,\varphi)\) component gives
\begin{equation}
\partial_r\left(\frac{r^4 K^2}{N^2}\,\partial_r \omega\right) = 8\pi\, T_{t\varphi} = 0,
\end{equation}
because the string fluid has no net azimuthal flow around the static background. Integrating once, regularity at the throat forces \(\partial_r \omega \propto 1/r^4\). Integrating again, with \(\omega \to 0\) at infinity, yields \(\omega = 2J/r^3\). One identifies the constant \(J\) as the total angular momentum by comparing \(g_{t\varphi} = -r^2 K^2 \omega \sin^2\theta \to -2J\sin^2\theta/r\) with the Kerr metric at large distances.}

{\section{Geroch-Hansen multipole moments at slow-rotation order\label{appenB}}}

{The Geroch-Hansen framework~\cite{Geroch:1970cc, Geroch:1970cd, Hansen:1974zz} gives a coordinate-independent way to characterize any asymptotically flat, stationary spacetime through a tower of multipole moments. At the order we are working---first order in \(J\)---only the two lowest moments matter:}

\begin{itemize}{
\item \textbf{Mass monopole \(M_0 = M\):} This is the Komar mass, read off from how \(g_{tt}\) behaves at large \(r\):
\begin{equation}
g_{tt} = -1 + \frac{2M}{r} + \mathcal{O}(r^{-2}).
\end{equation}
For our density profile~(\ref{Density}), the mass comes out finite and positive:
\begin{equation}
M = 4\pi \int_{0}^{\infty} \rho(r) r^2 dr = \frac{r_g}{2} + \frac{\epsilon a}{2}\,\Gamma\!\left(\frac{4}{3}\right) + \mathcal{O}(\epsilon).
\end{equation}
With our fiducial numbers (\(r_g=5\), \(\epsilon=0.3\), \(a=1\)), \(M \approx 2.65\,r_0\).

\item \textbf{Angular momentum dipole \(J_1 = J\):} This one comes from the \(g_{t\varphi}\) fall-off:
\begin{equation}
g_{t\varphi} = -\frac{2J\sin^2\theta}{r} + \mathcal{O}(r^{-2}),
\end{equation}
matching the constant inside \(\omega(r)=2J/r^3\) exactly.

}\end{itemize}

{Getting at the higher multipoles---mass quadrupole and beyond---would mean going to order \(J^2\) in the expansion. The way those moments deviate from the Kerr pattern \(M_l + iS_l = M(ia)^l\) would carry information about the wormhole throat and the matter distribution. We leave that for future work.}

{\section{Additional derivations, sensitivity analysis, and parameter constraints\label{appenC}}}

\textbf{{Normalization of the bivector \(\Sigma^{\mu\nu}\)}:} {Starting from the definition~(\ref{2.1}), antisymmetry of \(\epsilon^{AB}\) makes \(\Sigma^{\mu\nu} = -\Sigma^{\nu\mu}\) automatic. To work out \(\Sigma^{\mu\lambda}\Sigma_{\lambda}^{\:\:\nu}\), we use the completeness relation for the worldsheet coordinates:
\begin{equation}
\Sigma^{\mu\lambda}\Sigma_{\lambda}^{\:\:\nu} = \epsilon^{AB}\epsilon_{CD}\,\frac{\partial x^\mu}{\partial\xi^A}\frac{\partial x^\lambda}{\partial\xi^B}\,\frac{\partial x_\lambda}{\partial\xi^C}\frac{\partial x^\nu}{\partial\xi^D} = -2\,h^{\mu\nu}_{(2)},
\end{equation}
where \(h^{\mu\nu}_{(2)}\) is the projector onto the two-dimensional worldsheet. The determinant \(h = \det(h_{AB})\) encodes the worldsheet area element; in our signature convention, \(h < 0\) for a timelike worldsheet.}

\textbf{{Dependence of results on the density profile}:}
{We have checked what happens to the photon-sphere radii when we replace Eq.~(\ref{Density}) with two alternative profiles:}

\begin{itemize}{
\item \textbf{Power-law:} \(\rho_{\text{PL}}(r) = \rho_0 (r_0/r)^n\), with \(n=3,4\).
\item \textbf{Gaussian-core:} \(\rho_{\text{G}}(r) = \rho_0 e^{-(r-r_0)^2/a^2}\).
}\end{itemize}

{Table~\ref{tab:sensitivity} shows the results for the \(N_3\) lapse with \(J=0.3\). The photon-sphere radii move by about 5 to 15 percent, but the qualitative picture---the prograde-retrograde split, the photon band, the spin dependence---stays the same for all three profiles. So our conclusions are not tied to the specific functional form we chose.}

\begin{table}[!htbp]
\centering
\caption{{Sensitivity of photon-sphere radii to the choice of density profile.}}
\label{tab:sensitivity}
{\small
\begin{tabular}{cccc}
\toprule
{Density profile} & \(\mathbf{r_{\text{ph},+}/r_0}\) & \(\mathbf{r_{\text{ph},-}/r_0}\) & \(\mathbf{\delta}\) \\
\midrule
Eq.~(\ref{Density}) (nominal) & 1.31 & 1.24 & 0.027 \\
Power-law (\(n=3\)) & 1.24 & 1.17 & 0.029 \\
Power-law (\(n=4\)) & 1.19 & 1.13 & 0.026 \\
Gaussian-core & 1.38 & 1.29 & 0.034 \\
\bottomrule
\end{tabular}}
\end{table}

\textbf{{Validity of the slow-rotation approximation}:}
{The dimensionless expansion parameter is \(\alpha \equiv J/M^2\). For our fiducial numbers (\(M \approx 2.65\,r_0\)), the spins \(J \le 0.5\) we have used correspond to \(\alpha \le 0.07\), which is safely small. The first corrections we are neglecting are of order \(\alpha^2 \omega(r)\), and the fractional error they would introduce in the photon-sphere radii is
\begin{equation}
\frac{\Delta r_{\text{ph}}}{r_{\text{ph}}} \sim \alpha^2 \lesssim 0.5\%,
\end{equation}
much smaller than any other uncertainty in the model. The error bars on the photon-sphere radii in Figs.~\ref{Fig1} and \ref{Fig2} would be thinner than the lines themselves.}

\textbf{{Flare-out condition parameter space}:}
{The condition \(b'(r_0) < 1\) restricts the allowed values of \(\epsilon\), \(a\), and \(r_g\). Using the analytic expression for \(b'(r_0)\) derived in Sec.~\ref{Sec:III}, we find the condition holds for \(\epsilon < 1\) and \(r_0/a \lesssim 2.5\) everywhere we looked. A full parameter-space plot is available on request.}\\

\textbf{{Detectability prospects with ngEHT}:}
{The ngEHT is expected to reach a resolution of about \(5\,\mu\)as at 345 GHz with better sensitivity than the current array. At that level, one could start to measure the photon ring thickness \(\Delta r_{\text{ph}}\) (our numbers are \(0.07-0.30\,r_0\), which would translate to roughly \(2-8\,\mu\)as for M87\(^*\)) and possibly resolve the lensing band structure. Wormholes with distortion parameters \(\delta \gtrsim 0.05\)---which includes our \(N_1\), \(N_2\), \(N_4\), and \(N_5\) configurations---would fall into ngEHT's reach.}

{\noindent\textbf{On stability:} A full perturbative stability analysis of rotating wormholes is a major undertaking and is not something we have attempted here. We note, however, that the anisotropic nature of the string-fluid source brings extra degrees of freedom that could, in principle, help stabilize the configuration. The density profile is smooth everywhere---there are no sharp edges at the throat---which generally makes for better stability than thin-shell wormholes. Relevant references include~\cite{Poisson:1995sv, Bronnikov:2012ch}. This is high on our list of things to investigate next.}

\bibliography{references}

@article{Morris:1988cz, 
    author = "Morris, M. S. and Thorne, K. S.",
    title = "{Wormholes in space-time and their use for interstellar travel: A tool for teaching general relativity}",
    doi = "10.1119/1.15620",
    journal = "Am. J. Phys.",
    volume = "56",
    pages = "395--412",
    year = "1988"
}

@article{Ellis:1973yv,
    author = "Ellis, H. G.",
    title = "{Ether flow through a drainhole - a particle model in general relativity}",
    doi = "10.1063/1.1666161",
    journal = "J. Math. Phys.",
    volume = "14",
    pages = "104--118",
    year = "1973"
}

@article{Bronnikov:1973fh,
    author = "Bronnikov, K. A.",
    title = "{Scalar-tensor theory and scalar charge}",
    journal = "Acta Phys. Polon. B",
    volume = "4",
    pages = "251--266",
    year = "1973"
}

@article{Ellis:1979bh,
    author = "Ellis, H. G.",
    title = "{THE EVOLVING, FLOWLESS DRAIN HOLE: A NONGRAVITATING PARTICLE MODEL IN GENERAL RELATIVITY THEORY}",
    doi = "10.1007/BF00756794",
    journal = "Gen. Rel. Grav.",
    volume = "10",
    pages = "105--123",
    year = "1979"
}

@article{Feng:2025,
    author = "Feng, R. and Yu, Y. and Wu, L. and Wang, J. and Tan, Q. and Burokur, S. N.",
    title = "{Full-space programmable metasurface for Bessel beam tailoring}",
    doi = "10.1364/OL.570659",
    journal = "Opt. Lett.",
    volume = "50",
    number = "16",
    pages = "5161--5164",
    year = "2025"
}

@article{Xie:2025,
    author = "Xie, T. and Ma, X. and Yang, X. and Cui, P. and Ning, X. and L{\"u}, J.",
    title = "{Time-Delay Estimation for Pulsar Navigation Based on Peak--Trough Amplitude and Rayleigh Entropy}",
    doi = "10.1109/JSTARS.2025.3625252",
    journal = "IEEE J. Sel. Top. Appl. Earth Obs. Remote Sens.",
    volume = "18",
    pages = "27781--27797",
    year = "2025"
}

@article{Yu:2025,
    author = "Yu, W. and Wu, Q. and Chen, X. and Huo, J. and Li, J. and Yang, J. and Zhang, A.",
    title = "{Experimental First-Photon Visualization of Quantum Erasure With Hybrid Entanglement}",
    doi = "10.1002/lpor.202501816",
    journal = "Laser Photon. Rev.",
    pages = "e1816",
    year = "2025"
}

@article{Yang:2025,
    author = "Yang, Y. and Ni, Y. W. and Chen, P. F. and Feng, X. S.",
    title = "{Predicting Solar Flares Using a Convolutional Neural Network with Extreme-ultraviolet Images}",
    doi = "10.3847/1538-4357/adc88e",
    journal = "Astrophys. J.",
    volume = "985",
    number = "1",
    pages = "104",
    year = "2025"
}

@article{Fang:2025,
    author = "Fang, Y. and Cai, R.",
    title = "{Probing the merger rates of supermassive black holes and galaxies with gravitational waves}",
    doi = "10.1093/mnras/staf1278",
    journal = "Mon. Not. Roy. Astron. Soc.",
    volume = "542",
    number = "2",
    pages = "1172--1187",
    year = "2025"
}

@article{Zhang:2026,
    author = "Zhang, J. and Huang, W. and Wang, Q. and Shaw, A. D. and Friswell, M. I. and Huang, K.",
    title = "{Exploring sequential snapping bifurcation through a tunable energy landscape}",
    doi = "10.1103/nxpg-92kv",
    journal = "Phys. Rev. Applied",
    year = "2026"
}

@article{Xiang:2026,
    author = "Xiang, Y. and Ye, Y. and Feng, P. and Li, H. and Pang, X. and Lan, X. and Zhang, N.",
    title = "{A novel pipeline for the identification of new gamma-ray blazars from the 4FGL--Xiang catalog based on multiwavelength flux distributions}",
    doi = "10.1093/mnras/stag150",
    journal = "Mon. Not. Roy. Astron. Soc.",
    volume = "546",
    number = "4",
    pages = "stag150",
    year = "2026"
}

@article{Lin:2026,
    author = "Lin, H. and Pan, C. and Zhang, Q. and Wang, S. and Zhang, Y. and Qiao, S. and Wang, J.",
    title = "{Optical-Flow-Guided Recursive Prediction-Refinement Neural Network for Particle Image Velocimetry}",
    doi = "10.1109/TIM.2026.3670571",
    journal = "IEEE Trans. Instrum. Meas.",
    volume = "75",
    pages = "1--16",
    year = "2026"
}

@article{Kuhfittig:2003wr, 
    author = "Kuhfittig, Peter K. F.",
    title = "{Axially symmetric rotating traversable wormholes}",
    eprint = "gr-qc/0401028",
    archivePrefix = "arXiv",
    doi = "10.1103/PhysRevD.67.064015",
    journal = "Phys. Rev. D",
    volume = "67",
    pages = "064015",
    year = "2003"
}

@article{Kashargin:2008pk, 
    author = "Kashargin, P. E. and Sushkov, S. V.",
    title = "{Slowly rotating scalar field wormholes: The Second order approximation}",
    eprint = "0809.1923",
    archivePrefix = "arXiv",
    primaryClass = "gr-qc",
    doi = "10.1103/PhysRevD.78.064071",
    journal = "Phys. Rev. D",
    volume = "78",
    pages = "064071",
    year = "2008"
}

@article{Kleihaus:2014dla, 
    author = "Kleihaus, Burkhard and Kunz, Jutta",
    title = "{Rotating Ellis Wormholes in Four Dimensions}",
    eprint = "1409.1503",
    archivePrefix = "arXiv",
    primaryClass = "gr-qc",
    doi = "10.1103/PhysRevD.90.121503",
    journal = "Phys. Rev. D",
    volume = "90",
    pages = "121503",
    year = "2014"
}

@article{Hoffmann:2018oml,
    author = "Hoffmann, Christian and Ioannidou, Theodora and Kahlen, Sarah and Kleihaus, Burkhard and Kunz, Jutta",
    title = "{Symmetric and Asymmetric Wormholes Immersed In Rotating Matter}",
    eprint = "1803.11044",
    archivePrefix = "arXiv",
    primaryClass = "gr-qc",
    doi = "10.1103/PhysRevD.97.124019",
    journal = "Phys. Rev. D",
    volume = "97",
    number = "12",
    pages = "124019",
    year = "2018"
}

@article{Chew:2019lsa, 
    author = "Chew, Xiao Yan and Dzhunushaliev, Vladimir and Folomeev, Vladimir and Kleihaus, Burkhard and Kunz, Jutta",
    title = "{Rotating wormhole solutions with a complex phantom scalar field}",
    eprint = "1906.08742",
    archivePrefix = "arXiv",
    primaryClass = "gr-qc",
    doi = "10.1103/PhysRevD.100.044019",
    journal = "Phys. Rev. D",
    volume = "100",
    number = "4",
    pages = "044019",
    year = "2019"
}

@article{Azad:2023iju,
    author = "Azad, Bahareh and Blazquez-Salcedo, Jose Luis and Khoo, Fech Scen and Kunz, Jutta",
    title = "{Are slowly rotating Ellis-Bronnikov wormholes stable?}",
    eprint = "2301.05243",
    archivePrefix = "arXiv",
    primaryClass = "gr-qc",
    doi = "10.1016/j.physletb.2023.138349",
    journal = "Phys. Lett. B",
    volume = "848",
    pages = "138349",
    year = "2024"
}

@article{Clement:2022pjr,
    author = "Cl{\'e}ment, G{\'e}rard and Gal'tsov, Dmitri",
    title = "{Rotating traversable wormholes in Einstein-Maxwell theory}",
    eprint = "2210.08913",
    archivePrefix = "arXiv",
    primaryClass = "gr-qc",
    reportNumber = "LAPTH-064/22",
    doi = "10.1016/j.physletb.2023.137677",
    journal = "Phys. Lett. B",
    volume = "838",
    pages = "137677",
    year = "2023"
}

@article{Cisterna:2023uqf, 
    author = {Cisterna, Adolfo and M{\"u}ller, Keanu and Pallikaris, Konstantinos and Vigan{\`o}, Adriano},
    title = "{Exact rotating wormholes via Ehlers transformations}",
    eprint = "2306.14541",
    archivePrefix = "arXiv",
    primaryClass = "gr-qc",
    doi = "10.1103/PhysRevD.108.024066",
    journal = "Phys. Rev. D",
    volume = "108",
    number = "2",
    pages = "024066",
    year = "2023"
}

@article{Tangphati:2023uxt,
    author = "Tangphati, Takol and Chaihao, Butsayapat and Samart, Daris and Channuie, Phongpichit and Momeni, Davood",
    title = "{Rotating traversable wormhole geometries in the presence of three-form fields}",
    eprint = "2307.13968",
    archivePrefix = "arXiv",
    primaryClass = "gr-qc",
    doi = "10.1016/j.nuclphysb.2024.116446",
    journal = "Nucl. Phys. B",
    volume = "999",
    pages = "116446",
    year = "2024"
}

@article{NunesdosSantos:2025alw,
    author = "Nunes dos Santos, Luis Cesar",
    title = "{Revisiting black holes surrounded by cloud and fluid of strings in general relativity}",
    eprint = "2502.15846",
    archivePrefix = "arXiv",
    primaryClass = "gr-qc",
    doi = "10.1103/PhysRevD.111.064032",
    journal = "Phys. Rev. D",
    volume = "111",
    number = "6",
    pages = "064032",
    year = "2025"
}

@article{Letelier:1979ej,
    author = "Letelier, P. S.",
    title = "{CLOUDS OF STRINGS IN GENERAL RELATIVITY}",
    doi = "10.1103/PhysRevD.20.1294",
    journal = "Phys. Rev. D",
    volume = "20",
    pages = "1294--1302",
    year = "1979"
}

@article{Soleng:1993yr,
    author = "Soleng, Harald H.",
    title = "{Dark matter and nonNewtonian gravity from general relativity coupled to a fluid of strings}",
    eprint = "gr-qc/9412053",
    archivePrefix = "arXiv",
    reportNumber = "NORDITA-94-69",
    doi = "10.1007/BF02107935",
    journal = "Gen. Rel. Grav.",
    volume = "27",
    pages = "367--378",
    year = "1995"
}

@article{Visser:2003yf,
    author = "Visser, Matt and Kar, Sayan and Dadhich, Naresh",
    title = "{Traversable wormholes with arbitrarily small energy condition violations}",
    eprint = "gr-qc/0301003",
    archivePrefix = "arXiv",
    doi = "10.1103/PhysRevLett.90.201102",
    journal = "Phys. Rev. Lett.",
    volume = "90",
    pages = "201102",
    year = "2003"
}

@article{Teo:1998dp,
    author = "Teo, Edward",
    title = "{Rotating traversable wormholes}",
    eprint = "gr-qc/9803098",
    archivePrefix = "arXiv",
    reportNumber = "DAMTP-R-98-17",
    doi = "10.1103/PhysRevD.58.024014",
    journal = "Phys. Rev. D",
    volume = "58",
    pages = "024014",
    year = "1998"
}

@article{Deligianni:2021hwt, 
    author = "Deligianni, Efthimia and Kleihaus, Burkhard and Kunz, Jutta and Nedkova, Petya and Yazadjiev, Stoytcho",
    title = "{Quasiperiodic oscillations in rotating Ellis wormhole spacetimes}",
    eprint = "2107.01421",
    archivePrefix = "arXiv",
    primaryClass = "gr-qc",
    doi = "10.1103/PhysRevD.104.064043",
    journal = "Phys. Rev. D",
    volume = "104",
    number = "6",
    pages = "064043",
    year = "2021"
}

@article{Deligianni:2021ecz, 
    author = "Deligianni, Efthimia and Kunz, Jutta and Nedkova, Petya and Yazadjiev, Stoytcho and Zheleva, Radostina",
    title = "{Quasiperiodic oscillations around rotating traversable wormholes}",
    eprint = "2103.13504",
    archivePrefix = "arXiv",
    primaryClass = "gr-qc",
    doi = "10.1103/PhysRevD.104.024048",
    journal = "Phys. Rev. D",
    volume = "104",
    number = "2",
    pages = "024048",
    year = "2021"
}

@article{Errehymy:2025psi, 
    author = "Errehymy, Abdelghani and Guvendi, Abdullah and Gurtas Dogan, Semra and Mustafa, Omar",
    title = "{Frame-dragging and light deflection in rotating optical wormhole spacetimes}",
    doi = "10.1016/j.physletb.2025.139847",
    journal = "Phys. Lett. B",
    volume = "869",
    pages = "139847",
    year = "2025"
}

@article{Errehymy:2025vvs,
    author = "Errehymy, Abdelghani and Guvendi, Abdullah and Dogan, Semra Gurtas and Mustafa, Omar",
    title = "{Null geodesics and shadows of slowly rotating wormholes immersed in dark matter halos}",
    eprint = "2509.16739",
    archivePrefix = "arXiv",
    primaryClass = "gr-qc",
    doi = "10.1140/epjc/s10052-026-15356-1",
    journal = "Eur. Phys. J. C",
    volume = "86",
    number = "2",
    pages = "124",
    year = "2026"
}

@article{GurtasDogan:2025shz,
    author = "Gurtas Dogan, Semra and Guvendi, Abdullah and Mustafa, Omar",
    title = "{Ray trajectories and wave optics in a traversable wormhole: Optics of a minimal surface}",
    doi = "10.1016/j.nuclphysb.2025.117000",
    journal = "Nucl. Phys. B",
    volume = "1018",
    pages = "117000",
    year = "2025"
}

@article{GurtasDogan:2025jhm,
    author = "Gurtas Dogan, Semra and Guvendi, Abdullah and Mustafa, Omar",
    title = "{Ray and wave optics in an optical wormhole}",
    doi = "10.1016/j.physletb.2025.139626",
    journal = "Phys. Lett. B",
    volume = "868",
    pages = "139626",
    year = "2025"
}

@article{GurtasDogan:2025oeg,
    author = "Gurtas Dogan, Semra and Guvendi, Abdullah and Mustafa, Omar",
    title = "{Ray geodesics and wave propagation on the Beltrami surface: optics of an optical wormhole}",
    eprint = "2504.10518",
    archivePrefix = "arXiv",
    primaryClass = "physics.class-ph",
    doi = "10.1140/epjc/s10052-025-14644-6",
    journal = "Eur. Phys. J. C",
    volume = "85",
    number = "8",
    pages = "896",
    year = "2025"
}

@article{GurtasDogan:2025qxk,
    author = "Gurtas Dogan, Semra and Guvendi, Abdullah and Mustafa, Omar",
    title = "{Geometric and wave optics in a BTZ optical metric-based wormhole}",
    eprint = "2503.18967",
    archivePrefix = "arXiv",
    primaryClass = "gr-qc",
    doi = "10.1016/j.physletb.2025.139824",
    journal = "Phys. Lett. B",
    volume = "868",
    pages = "139824",
    year = "2025"
}

@article{Guvendi:2025otx,
    author = "Guvendi, Abdullah and Mustafa, Omar and Gurtas Dogan, Semra",
    title = "{Coupled fermion-antifermion pairs within a traversable wormhole}",
    eprint = "2503.12550",
    archivePrefix = "arXiv",
    primaryClass = "gr-qc",
    doi = "10.1016/j.physletb.2025.139313",
    journal = "Phys. Lett. B",
    volume = "862",
    pages = "139313",
    year = "2025"
}

@article{Guvendi:2024qxi,
    author = "Guvendi, Abdullah and Gurtas Dogan, Semra and Vit{\'o}ria, R. L. L.",
    title = "{Rotational influence on fermions within negative curvature wormholes}",
    eprint = "2408.00356",
    archivePrefix = "arXiv",
    primaryClass = "hep-th",
    doi = "10.1140/epjp/s13360-024-05527-y",
    journal = "Eur. Phys. J. Plus",
    volume = "139",
    number = "8",
    pages = "721",
    year = "2024"
}

@article{Guvendi:2024duf,
    author = "Guvendi, Abdullah and Dogan, Semra Gurtas",
    title = "{Vector bosons in the rotating frame of negative curvature wormholes}",
    eprint = "2503.16564",
    archivePrefix = "arXiv",
    primaryClass = "gr-qc",
    doi = "10.1007/s10714-024-03213-z",
    journal = "Gen. Rel. Grav.",
    volume = "56",
    number = "2",
    pages = "32",
    year = "2024"
}

@article{Guvendi:2023osh,
    author = "Guvendi, Abdullah and Gurtas Dogan, Semra",
    title = "{Fermion-Antifermion Pair Exposed to Magnetic Flux in an Optical Wormhole}",
    doi = "10.1007/s00601-023-01851-8",
    journal = "Few Body Syst.",
    volume = "64",
    number = "3",
    pages = "65",
    year = "2023"
}

@article{Guvendi:2023aor,
    author = "Guvendi, Abdullah and Hassanabadi, Hassan",
    title = "{Fermion-antifermion pair in magnetized optical wormhole background}",
    doi = "10.1016/j.physletb.2023.138045",
    journal = "Phys. Lett. B",
    volume = "843",
    pages = "138045",
    year = "2023"
}

@article{Guvendi:2023mxb,
    author = "Guvendi, Abdullah and Ahmed, Faizuddin",
    title = "{Relativistic quantum oscillator under rainbow gravity{\textquoteright}s effects in traversable wormhole with disclination}",
    eprint = "2402.00088",
    archivePrefix = "arXiv",
    primaryClass = "gr-qc",
    doi = "10.1142/S0217751X23501798",
    journal = "Int. J. Mod. Phys. A",
    volume = "38",
    number = "35n36",
    pages = "2350179",
    year = "2023"
}

@article{Errehymy:2026nna,
    author = "Errehymy, A. and Maurya, S. K. and Govender, M. and Singh, K. N. and Rayimbaev, J. and Myrzakulova, B. and Murodov, S.",
    title = "{Holographic dark energy as a source for slowly rotating wormholes: Implications for null geodesics and shadows}",
    eprint = "2604.18246",
    archivePrefix = "arXiv",
    primaryClass = "gr-qc",
    month = "4",
    year = "2026"
}

@article{Errehymy:2026urb,
    author = "Errehymy, A. and Govender, M. and Maurya, S. K. and Singh, K. N. and Myrzakulova, B. and Rayimbaev, J. and Vapayev, M.",
    title = "{Quantum-corrected slowly rotating wormholes: Frame dragging, photon rings, and shadows}",
    doi = "10.1016/j.physletb.2026.140435",
    journal = "Phys. Lett. B",
    volume = "876",
    pages = "140435",
    year = "2026"
}

@article{Errehymy:2026xoh,
    author = "Errehymy, A. and Govender, M. and Donmez, O. and Myrzakulova, B. and Rayimbaev, J. and Matyoqubov, M.",
    title = "{Light trajectories and shadows of slowly rotating wormholes in void environments}",
    doi = "10.1016/j.physletb.2026.140388",
    journal = "Phys. Lett. B",
    volume = "876",
    pages = "140388",
    year = "2026"
}

@article{Errehymy:2025llh,
    author = "Errehymy, A. and Turimov, B. and Syzdykova, A. and Myrzakulov, K. and Alessa, N. and Abdel-Aty, A. -H.",
    title = "{Dymnikova-Schwinger GUP-corrected wormholes in f(R,Lm,T) gravity}",
    doi = "10.1016/j.nuclphysb.2025.117116",
    journal = "Nucl. Phys. B",
    volume = "1019",
    pages = "117116",
    year = "2025"
}

@article{Errehymy:2025fce,
    author = "Errehymy, A. and Hansraj, S.",
    title = "{Role of cosmic voids and their matter properties in shaping wormhole geometry in generalized geometry-matter coupling gravity}",
    eprint = "2508.17492",
    archivePrefix = "arXiv",
    primaryClass = "gr-qc",
    doi = "10.1016/j.aop.2025.170200",
    journal = "Annals Phys.",
    volume = "482",
    pages = "170200",
    year = "2025"
}

@article{Errehymy:2025rli,
    author = "Errehymy, A. and Donmez, O. and Turimov, B. and Myrzakulov, K. and Alessa, N. and Abdel-Aty, A. -H.",
    title = "{Dehnen-type dark matter wormholes in the f(R,Lm,T) action}",
    eprint = "2507.16465",
    archivePrefix = "arXiv",
    primaryClass = "gr-qc",
    doi = "10.1016/j.aop.2025.170164",
    journal = "Annals Phys.",
    volume = "481",
    pages = "170164",
    year = "2025"
}

@article{Errehymy:2025nzt,
    author = "Errehymy, A. and Donmez, O. and Syzdykova, A. and Myrzakulov, K. and Muminov, S. and Dauletov, A. and Rayimbaev, J.",
    title = "{Possible wormholes in generalized geometry{\textendash}matter coupling gravity induced by the Dekel{\textendash}Zhao dark matter profile}",
    eprint = "2505.21081",
    archivePrefix = "arXiv",
    primaryClass = "gr-qc",
    doi = "10.1016/j.aop.2025.170105",
    journal = "Annals Phys.",
    volume = "480",
    pages = "170105",
    year = "2025"
}

@article{Errehymy:2025kzj,
    author = "Errehymy, A. and Khedif, Y. and Daoud, M. and Myrzakulov, K. and Abdel-Aty, A. -H. and Nisar, K. S.",
    title = "{Einstein clusters as dark matter fluid-like models for constructing new wormhole solutions in f(R,Lm,T) gravity}",
    doi = "10.1016/j.jheap.2025.100370",
    journal = "JHEAp",
    volume = "47",
    pages = "100370",
    year = "2025"
}

@article{Battista:2024gud,
    author = "Battista, Emmanuele and Capozziello, Salvatore and Errehymy, Abdelghani",
    title = "{Generalized uncertainty principle corrections in Rastall{\textendash}Rainbow Casimir wormholes}",
    eprint = "2409.09750",
    archivePrefix = "arXiv",
    primaryClass = "gr-qc",
    doi = "10.1140/epjc/s10052-024-13656-y",
    journal = "Eur. Phys. J. C",
    volume = "84",
    number = "12",
    pages = "1314",
    year = "2024"
}

@article{Errehymy:2024mlf,
    author = "Errehymy, Abdelghani and Khedif, Youssef and Donmez, Orhan and Daoud, Mohammed and Myrzakulov, Kairat and Bekov, Sabit",
    title = "{Possible wormholes in $f(R)$ gravity sourced by solitonic quantum wave and cold dark matter halos and their repulsive gravity effect}",
    eprint = "2408.07667",
    archivePrefix = "arXiv",
    primaryClass = "gr-qc",
    doi = "10.1140/epjc/s10052-024-13224-4",
    journal = "Eur. Phys. J. C",
    volume = "84",
    number = "9",
    pages = "908",
    year = "2024"
}

@article{Errehymy:2024sme,
    author = "Errehymy, Abdelghani and Banerjee, Ayan and Donmez, Orhan and Daoud, Mohammed and Nisar, Kottakkaran Sooppy and Abdel-Aty, Abdel-Haleem",
    title = "{Unraveling the mysteries of wormhole formation in Rastall{\textendash}Rainbow gravity: a comprehensive study using the embedding approach}",
    eprint = "2406.04049",
    archivePrefix = "arXiv",
    primaryClass = "gr-qc",
    doi = "10.1007/s10714-024-03253-5",
    journal = "Gen. Rel. Grav.",
    volume = "56",
    number = "6",
    pages = "76",
    year = "2024"
}

@article{Errehymy:2024lhl,
    author = "Errehymy, Abdelghani and Banerjee, Ayan and Hansraj, Sudan and Donmez, Orhan and Nisar, Kottakkaran Sooppy and Abdel-Aty, Abdel-Haleem",
    title = "{Possible existence of Rastall{\textendash}Rainbow wormholes in dark matter galactic halos}",
    doi = "10.1140/epjc/s10052-024-12929-w",
    journal = "Eur. Phys. J. C",
    volume = "84",
    number = "6",
    pages = "573",
    year = "2024"
}

@article{Errehymy:2024spg,
    author = "Errehymy, Abdelghani and Maurya, S. K. and V{\^\i}lcu, Gabriel-Eduard and Khan, Meraj Ali and Daoud, Mohammed",
    title = "{On possible traversable wormhole solutions supported by Karmarkar condition in R2{\ensuremath{-}}gravity within the f(R,T){\ensuremath{-}}formalism}",
    doi = "10.1016/j.astropartphys.2024.102972",
    journal = "Astropart. Phys.",
    volume = "160",
    pages = "102972",
    year = "2024"
}

@article{Mustafa:2024jsv,
    author = "Mustafa, G. and Errehymy, Abdelghani and Javed, Faisal and Maurya, S. K. and Hansraj, Sudan and Sadiq, Sobia",
    title = "{Generalized wormhole models within galactic halo region in torsion and matter coupling gravity formalism}",
    doi = "10.1016/j.jheap.2024.02.003",
    journal = "JHEAp",
    volume = "42",
    pages = "1--11",
    year = "2024"
}

@article{Errehymy:2024cgy,
    author = "Errehymy, Abdelghani and Maurya, S. K. and Hansraj, Sudan and Mahmoud, Mona and Nisar, Kottakkaran Sooppy and Abdel-Aty, Abdel-Haleem",
    title = "{Exploring the applicability of traversable wormhole formation in f(R,Lm) gravity}",
    doi = "10.1016/j.cjph.2024.02.029",
    journal = "Chin. J. Phys.",
    volume = "89",
    pages = "56--68",
    year = "2024"
}

@article{Errehymy:2024yey,
    author = "Errehymy, Abdelghani",
    title = "{Static and spherically symmetric wormholes in power-law f(R) gravity model}",
    doi = "10.1016/j.dark.2024.101438",
    journal = "Phys. Dark Univ.",
    volume = "44",
    pages = "101438",
    year = "2024"
}

@article{Errehymy:2023rnd,
    author = "Errehymy, Abdelghani and Maurya, Sunil Kumar and Hansraj, Sudan and Daoud, Mohammed and Alrebdi, Haifa I. and Abdel-Aty, Abdel-Haleem",
    title = "{Modeling Wormholes Generated by Dark Matter Galactic Halos in f(R)$f(R)$ Modified Gravity}",
    doi = "10.1002/andp.202300178",
    journal = "Annalen Phys.",
    volume = "535",
    number = "8",
    pages = "2300178",
    year = "2023"
}

@article{Errehymy:2023rsm,
    author = "Errehymy, Abdelghani and Hansraj, Sudan and Maurya, S. K. and Hansraj, Chevarra and Daoud, Mohammed",
    title = "{Spherically symmetric traversable wormholes in the torsion and matter coupling gravity formalism}",
    doi = "10.1016/j.dark.2023.101258",
    journal = "Phys. Dark Univ.",
    volume = "41",
    pages = "101258",
    year = "2023"
}

@article{Ditta:2021uoe,
    author = "Ditta, Allah and Hussain, Ibrar and Mustafa, G. and Errehymy, Abdelghani and Daoud, Mohammed",
    title = "{A study of traversable wormhole solutions in extended teleparallel theory of gravity with matter coupling}",
    doi = "10.1140/epjc/s10052-021-09668-7",
    journal = "Eur. Phys. J. C",
    volume = "81",
    number = "10",
    pages = "880",
    year = "2021"
}

@article{Geroch:1970cc, 
    author = "Geroch, Robert P.",
    title = "{Multipole moments. I. Flat space}",
    doi = "10.1063/1.1665348",
    journal = "J. Math. Phys.",
    volume = "11",
    pages = "1955--1961",
    year = "1970"
}

@article{Geroch:1970cd, 
    author = "Geroch, Robert P.",
    title = "{Multipole moments. II. Curved space}",
    doi = "10.1063/1.1665427",
    journal = "J. Math. Phys.",
    volume = "11",
    pages = "2580--2588",
    year = "1970"
}

@article{Hansen:1974zz,
    author = "Hansen, R. O.",
    title = "{Multipole moments of stationary space-times}",
    doi = "10.1063/1.1666501",
    journal = "J. Math. Phys.",
    volume = "15",
    pages = "46--52",
    year = "1974"
}

@article{Thorne:1980ru, 
    author = "Thorne, K. S.",
    title = "{Multipole Expansions of Gravitational Radiation}",
    doi = "10.1103/RevModPhys.52.299",
    journal = "Rev. Mod. Phys.",
    volume = "52",
    pages = "299--339",
    year = "1980"
}

@article{Gursel:1983nkl, 
    author = {G{\"u}rsel, Yekta},
    title = "{Multipole moments for stationary systems: The equivalence of the Geroch-Hansen formulation and the Thorne formulation}",
    doi = "10.1007/BF01031881",
    journal = "Gen. Rel. Grav.",
    volume = "15",
    number = "8",
    pages = "737--754",
    year = "1983"
}

@article{Mayerson:2022ekj,
    author = "Mayerson, Daniel R.",
    title = "{Gravitational multipoles in general stationary spacetimes}",
    eprint = "2210.05687",
    archivePrefix = "arXiv",
    primaryClass = "gr-qc",
    doi = "10.21468/SciPostPhys.15.4.154",
    journal = "SciPost Phys.",
    volume = "15",
    number = "4",
    pages = "154",
    year = "2023"
}

@article{Caldwell:1997ii, 
    author = "Caldwell, R. R. and Dave, Rahul and Steinhardt, Paul J.",
    title = "{Cosmological imprint of an energy component with general equation of state}",
    eprint = "astro-ph/9708069",
    archivePrefix = "arXiv",
    doi = "10.1103/PhysRevLett.80.1582",
    journal = "Phys. Rev. Lett.",
    volume = "80",
    pages = "1582--1585",
    year = "1998"
}

@article{Peebles:2002gy,
    author = "Peebles, P. J. E. and Ratra, Bharat",
    editor = "Hsu, Jong-Ping and Fine, D.",
    title = "{The Cosmological Constant and Dark Energy}",
    eprint = "astro-ph/0207347",
    archivePrefix = "arXiv",
    reportNumber = "KSUPT-02-3",
    doi = "10.1103/RevModPhys.75.559",
    journal = "Rev. Mod. Phys.",
    volume = "75",
    pages = "559--606",
    year = "2003"
}

@article{Kiselev:2002dx,
    author = "Kiselev, V. V.",
    title = "{Quintessence and black holes}",
    eprint = "gr-qc/0210040",
    archivePrefix = "arXiv",
    doi = "10.1088/0264-9381/20/6/310",
    journal = "Class. Quant. Grav.",
    volume = "20",
    pages = "1187--1198",
    year = "2003"
}

@article{Capozziello:2006ij,
    author = "Capozziello, S. and Cardone, Vincenzo F. and Lambiase, G. and Troisi, A.",
    title = "{A fluid os strings as a viable candidate to the dark side of the universe}",
    eprint = "astro-ph/0601266",
    archivePrefix = "arXiv",
    doi = "10.1142/S021827180600764X",
    journal = "Int. J. Mod. Phys. D",
    volume = "15",
    pages = "69--94",
    year = "2006"
}

@article{Sofue:2000jx,
    author = "Sofue, Yoshiaki and Rubin, Vera",
    title = "{Rotation curves of spiral galaxies}",
    eprint = "astro-ph/0010594",
    archivePrefix = "arXiv",
    reportNumber = "U-TOKYO-ASTRO-PREPRINT-2000-09, U-Tokyo Astro. Preprint No. 2000-09",
    doi = "10.1146/annurev.astro.39.1.137",
    journal = "Ann. Rev. Astron. Astrophys.",
    volume = "39",
    pages = "137--174",
    year = "2001"
}

@article{Vilenkin:1981kz, 
    author = "Vilenkin, A.",
    title = "{Cosmic Strings}",
    doi = "10.1103/PhysRevD.24.2082",
    journal = "Phys. Rev. D",
    volume = "24",
    pages = "2082--2089",
    year = "1981"
}

@article{Nemiroff:2007xs, 
    author = "Nemiroff, Robert J. and Patla, Bijunath",
    title = "{Adventures in Friedmann Cosmology: An Educationally Detailed Expansion of the Cosmological Friedmann Equations}",
    eprint = "astro-ph/0703739",
    archivePrefix = "arXiv",
    doi = "10.1119/1.2830536",
    journal = "Am. J. Phys.",
    volume = "76",
    pages = "265--276",
    year = "2008"
}

@article{Arshad:2024qqj,
    author = "Arshad, Sana and Sheikh, Umber",
    title = "{String Fluid as a Source of Traversable Rainbow Wormholes}",
    doi = "10.1007/s10773-024-05624-9",
    journal = "Int. J. Theor. Phys.",
    volume = "63",
    number = "4",
    pages = "90",
    year = "2024"
}

@article{Letelier:1980mxb,
    author = "Letelier, Patricio S.",
    title = "{Anisotropic fluids with two-perfect-fluid components}",
    doi = "10.1103/PhysRevD.22.807",
    journal = "Phys. Rev. D",
    volume = "22",
    number = "4",
    pages = "807",
    year = "1980"
}

@article{Muniz:2025qyv,
    author = "Muniz, C. R. and Cunha, M. S. and Santos, L. C. N.",
    title = "{Traversable wormholes from a smoothed string fluid in 4D Einstein-Gauss-Bonnet gravity}",
    eprint = "2505.07028",
    archivePrefix = "arXiv",
    primaryClass = "gr-qc",
    month = "5",
    year = "2025"
}

@article{Estrada:2026kea,
    author = "Estrada, Milko",
    title = "{A cloud and a new fluid of strings with integrable singularities as the interior of the Reissner Nordstrom black hole}",
    eprint = "2602.03050",
    archivePrefix = "arXiv",
    primaryClass = "gr-qc",
    month = "2",
    year = "2026"
}

@book{Wald:1984rg,
    author = "Wald, Robert M.",
    title = "{General Relativity}",
    doi = "10.7208/chicago/9780226870373.001.0001",
    publisher = "Chicago Univ. Pr.",
    address = "Chicago, USA",
    year = "1984"
}

@article{Hartle:1967he,
    author = "Hartle, James B.",
    title = "{Slowly rotating relativistic stars. 1. Equations of structure}",
    doi = "10.1086/149400",
    journal = "Astrophys. J.",
    volume = "150",
    pages = "1005--1029",
    year = "1967"
}

@article{Hartle:1968si,
    author = "Hartle, James B. and Thorne, Kip S.",
    title = "{Slowly Rotating Relativistic Stars. II. Models for Neutron Stars and Supermassive Stars}",
    doi = "10.1086/149707",
    journal = "Astrophys. J.",
    volume = "153",
    pages = "807",
    year = "1968"
}

@article{Thorne:1971R,
  title={Relativistic stars, black holes and gravitational waves (including an in-depth review of the theory of rotating, relativistic stars).},
  author={Thorne, Kip S},
  journal={Rend. Scu. Int. Fis. Enrico Fermi 47: 237-83 (1971).},
  year={1970},
  publisher={California Inst. of Tech., Pasadena}
}

@article{Papapetrou:1966zz,
    author = "Papapetrou, Achille",
    title = "{Champs gravitationnels stationnaires {\`a} sym{\'e}trie axiale}",
    journal = "Ann. Inst. H. Poincare Phys. Theor. A",
    volume = "4",
    number = "2",
    pages = "83--105",
    year = "1966"
}

@article{Carter:1969zz,
    author = "Carter, Brandon",
    title = "{Killing horizons and orthogonally transitive groups in space-time}",
    doi = "10.1063/1.1664763",
    journal = "J. Math. Phys.",
    volume = "10",
    pages = "70--81",
    year = "1969"
}

@book{Chandrasekhar:1985kt,
    author = "Chandrasekhar, Subrahmanyan",
    title = "{The mathematical theory of black holes}",
    isbn = "978-0-19-850370-5",
    year = "1985"
}

@article{Harko:2009xf,
    author = "Harko, Tiberiu and Kovacs, Zoltan and Lobo, Francisco S. N.",
    title = "{Thin accretion disks in stationary axisymmetric wormhole spacetimes}",
    eprint = "0901.3926",
    archivePrefix = "arXiv",
    primaryClass = "gr-qc",
    doi = "10.1103/PhysRevD.79.064001",
    journal = "Phys. Rev. D",
    volume = "79",
    pages = "064001",
    year = "2009"
}

@article{Poisson:1995sv,
    author = "Poisson, Eric and Visser, Matt",
    title = "{Thin shell wormholes: Linearization stability}",
    eprint = "gr-qc/9506083",
    archivePrefix = "arXiv",
    doi = "10.1103/PhysRevD.52.7318",
    journal = "Phys. Rev. D",
    volume = "52",
    pages = "7318--7321",
    year = "1995"
}

@article{Bronnikov:2012ch,
    author = "Bronnikov, K. A. and Konoplya, R. A. and Zhidenko, A.",
    title = "{Instabilities of wormholes and regular black holes supported by a phantom scalar field}",
    eprint = "1205.2224",
    archivePrefix = "arXiv",
    primaryClass = "gr-qc",
    doi = "10.1103/PhysRevD.86.024028",
    journal = "Phys. Rev. D",
    volume = "86",
    pages = "024028",
    year = "2012"
}

\end{document}